\documentstyle[11pt,aaspp4]{article}
\singlespace

\def\uhz{$\mu\rm{Hz}$ }

\def\m*{M_*}

\lefthead{Kawaler, Sekii, \& Gough}
\righthead{Differential Rotation in White Dwarfs}

\begin{document}

\title{Prospects for Measuring Differential Rotation in White Dwarfs Through
Asteroseismology}

\author{Steven D. Kawaler \altaffilmark{1,2},
        Takashi Sekii \altaffilmark{2} 
   \and Douglas Gough \altaffilmark{2,3}}

\altaffiltext{1}{Department of Physics and Astronomy, 
                Iowa State University, Ames, IA 50011 USA; email:
                 sdk@iastate.edu}
\altaffiltext{2}{Institute of Astronomy, University of Cambridge, Madingley
                 Road, Cambridge CB3 0HA, England}
\altaffiltext{3}{Department of Applied Mathematics and Theoretical Physics,
                 University of Cambridge, Silver Street, Cambridge CB3 9EW, 
                 England}

\begin{abstract}
We examine the potential of asteroseismology for exploring the internal
rotation of white dwarf stars.  Data from global observing campaigns have
revealed a wealth of frequencies, some of which show the signature of 
rotational splitting.  Tools developed for helioseismology to use many
solar p-mode frequencies for inversion of the rotation rate with depth are
adapted to the case of more limited numbers of modes of low degree.  We find
that the small number of available modes in white dwarfs, coupled with the
similarity between the rotational-splitting kernels of the modes, renders
direct inversion unstable.
Accordingly, we adopt what we consider to be plausible functional forms for
the differential rotation profile; this is sufficiently restrictive to enable
us to carry out a useful calibration.  We show examples of this technique for
PG~1159 stars and pulsating DB white dwarfs.  Published frequency splittings
for white dwarfs are currently not accurate enough for meaningful inversions;
reanalysis of existing data can provide splittings of sufficient accuracy
when the frequencies of individual peaks are extracted via least-squares
fitting or multipeak decompositions.  We find that when mode trapping is
evident in the period spacing of g modes, the measured splittings can
constrain ${\rm d}\Omega/{\rm d}r$.
\end{abstract} 


\section{Introduction}

In recent years, the study of the late evolutionary stages of low- and 
intermediate-mass stars has benefited significantly from the 
investigation of pulsating white dwarfs.  There are three classes of 
pulsating white dwarf stars: the pulsating DA white dwarfs (ZZ Ceti 
stars), the pulsating DB white dwarfs (i.e.  GD~358), and the hot, C/O 
rich GW Vir (or pulsating PG~1159) stars.  Through asteroseismological 
analysis of their pulsations, we have learned about their global 
properties (mass, luminosity, effective temperature, distance, 
rotation periods, ...) and their internal properties (compositional 
stratification, cooling rates, magnetic field strength, ...).  
Examples of the best-studied pulsating white dwarfs include 
PG~1159-035 (Winget et al.  1991, hereafter WWETPG, Kawaler \& Bradley 
1994), the pulsating DB white dwarf GD~358 (Winget et al.  1994, 
Bradley \& Winget 1994), and the ZZ Ceti star G29-38 (Kleinman et al.  
1994, Kleinman 1998); see also the reviews contained in proceedings of 
the Whole Earth Telescope Workshops (Meistas \& Solheim 1996, Meistas 
\& Moskalik 1998).  As the observational situation has improved, most 
notably with the continuing work of the Whole Earth Telescope (WET; 
Nather et al.  1990), there now exists the possibility for even more 
detailed probing of these stars.  For example, Winget et al.  (1994, 
hereafter WWETDB) report that the power spectrum of GD~358 shows 
evidence for a variation of rotation with depth, and of a modest 
magnetic field.  In this paper, we explore the possibilities of using 
the measured pulsation frequencies to constrain internal rotation 
profiles.

Traditional determination of rotation rates in stars involve measuring
rotational broadening of spectral lines.  In white dwarfs, the natural
broadening mechanisms render such rotational broadening measurements
insensitive to rotation periods longer than a few hours.  In some cases,
narrow Balmer line cores can be produced by NLTE effects (see Section 1.2
below) but ambiguity between NLTE subtleties and rotational broadening
complicates their interpretation (i.e. Koester et al. 1998).  Fortunately, 
one of the properties that can be provided by pulsation observations is the
rotation of the star.  A frequently used relation between the observed
frequency splitting and the angular velocity $\Omega$ of the star, valid when
$\Omega$ is constant (and small compared with $2\pi\nu_{nl}$), is
\begin{equation}
m\delta\nu_{nl}\equiv\nu_{nlm} - \nu_{n l 0} = \frac{m}{P_{\rm rot}}\beta_{nl}~,
\end{equation}
where $\nu_{nl0}$ is the cyclic frequency of oscillation of a mode of order
$n$ and degree $l$ in the absence of rotation, $\nu_{n l m}$ is the observed
frequency for a given multiplet component with azimuthal quantum number $m$,
and ${P_{\rm rot}}$ is the rotation period of the star. The quantity
$\beta_{nl}$ is a function of the oscillation eigenfunction of the
corresponding nonrotating stars, and is independent of $m$; for g modes of
white dwarf stars, a useful approximation, good to about 10 per cent and
frequently much better, is
\begin{equation}
\beta_{nl} \approx 1-\frac{1}{l(l+1)}
\end{equation}
(Brickhill 1975).  Another form for the rotation splitting constant is
$C_{nl}$, where $C_{nl}=1-\beta_{nl}$.

Expression (1) assumes that the angular velocity $\Omega$ is constant
throughout the star.  However, there is no reason {\it a priori} to believe 
that white dwarfs rotate uniformly.  If the putative variation of the angular
velocity with depth is taken into account, the resulting rotational 
splitting is
\begin{equation}
\delta \nu_{nl} = 2 \pi I^{-1} \int_0^R 
       \Omega(r) \left\{\xi^2 +2\xi\eta+[1-\frac{1}{l(l+1)}]\eta^2\right\} 
       \rho r^2 {\rm d}r,
\end{equation}
where $\xi$ and $\eta$ represent the amplitudes of the vertical and
horizontal displacement eigenfunctions, ignoring the perturbation due to
rotation. The normalization factor $I$ is the inertia of the mode. It is
given by
\begin{equation}
I= \int_0^R [\xi^2+\eta^2] \rho r^2 {\rm d}r.
\end{equation}
For typical observable g modes in white dwarf stars, $n \gg l$ and therefore
$\eta \gg \xi$; it is in the limit $l/n\rightarrow 0$ that the rotational
coefficient $\beta_{nl}$ for uniform rotation reaches the asymptotic value
given by equation (2).  As in equation (1), expression (3) for the mean
splitting $\delta\nu_{nl}$ is also independent of $m$.

The rotational splitting formula (3) represents a weighted average of the
rotation rate with depth.  A single splitting therefore provides us with only
some average rotation rate, weighted by
the displacement eigenfunction of the mode in question.  It is
convenient to rewrite equation (3) as follows:
\begin{equation}
\delta \nu_{n l} = 2 \pi \int_0^R K_{n l}(r) \Omega(r) {\rm d}r~,
\end{equation}
where the data kernel (weighting function) $K_{n l}(r)$ depends on the
rotationally unperturbed eigenfunction for the mode specified by $n$ and $l$,
and is a function of $r$ alone.  Note that the normalization of $K_{nl}$ is
such that the integral of $K_{nl}$ is $\beta_{nl}$, which for most of the
modes considered in this paper is approximately 0.5.  The rotation kernels
are available from the pulsation analysis of a representative equilibrium
stellar model.  Constraints on possible stellar models can be applied from
the observed frequencies of oscillation, and in principle can limit the
possibilities adequately, allowing us then to interpret the observed
splitting in somewhat more detail than identifying some ``average rotation
rate of the star''.

\subsection{Towards determining differential rotation}

When several rotational splittings have been measured in a star, the 
appropriate kernels for the modes observed can be used to constrain 
the depth dependence of the rotation rate.  Such constraints can be 
obtained by two procedures: the {\em forward calculation} and {\em 
inversion}.  In the forward procedure, the observed splittings are 
reproduced using the best available stellar model by adjusting the 
model internal rotation rate (somehow) to match the observed frequency 
splittings.  However, if a function $\Omega(r)$ can be found to 
produce a good match, then there are many --- formally infinitely many 
--- other functions $\Omega(r)$ that can do so too.  Moreover, it does 
not guarantee a straightforward analysis of the effects of 
observational errors, although Monte Carlo methods can easily be used.  
The forward calculation is a useful elementary way to explore the 
property of classes of `solutions' (such as power laws, flat rotation 
curves, etc.), and is relatively fast to compute.  In most analyses 
that have been carried out to date, rotational splittings are 
converted into rotation rates by assuming a flat rotation curve and 
using equation (1) or its equivalent.

The object of inversion is to `solve' equations (5), using all the observed
rotational splittings and kernels from an appropriate stellar model, to
obtain values of $\Omega$ in different regions of the stellar interior.
Evidently, with only a finite number of data, no unique function $\Omega(r)$
can be determined. Consequently additional constraints must be applied to
render the solution unique.  A natural constraint imposes smoothness, since
by their nature typical kernel sets cannot isolate small-scale structure.

Rotational inversion procedures have been explored extensively in the 
context of helioseismology, and have yielded quite detailed mapping of 
the sun's angular velocity as a function of depth and latitude.  We 
refer the reader to review by Gough \& Thompson (1991), and references 
therein, for the details.  Here we point out simply that the 
procedures fall basically into two categories.  In the first a class 
of acceptable (typically smooth) functions $\Omega(r)$ satisfying 
equation (5) (within the observational uncertainty) is sought.  One is 
chosen that is considered to be best, based on a criterion that 
balances satisfying the data and satisfying the externally imposed 
acceptability criterion (such as smoothness).  This method emphasizes 
reproducing the data, and is logically just a formalization of the 
search in the forward method mentioned above.  the other method is 
directed towards addressing actual properties of the rotation, rather 
than to satisfying the data.  It seeks linear combinations of data 
that yield equations of the type (5) but with more easily 
interpretable kernels (such as kernels that are highly localized in 
space and are nowhere (large and) negative.  The outcome of this 
method is simpler to interpret, since the results are local averages 
over known regions.  However, it is not always possible to construct 
an averaging kernel with the desired properties unless the data 
kernels associated with the data are sufficiently diverse that 
localized information is unambiguously contained in the data.  That is 
not the case of the data we discuss in this paper.  It is therefore 
prudent to adopt a method from the first category.  Helioseismic data 
contain the signals of millions of frequencies, over a range of degree 
up to and above $l=1000$.  For white dwarfs, however, $l$ is typically 
1, and occasionally 2.  Stars considered to be ``rich'' pulsators show 
20 separate multiplets.  Thus the prospect for detailed rotational 
inversion for white dwarfs is limited.  We note that Goupil et al.  
(1996) explored some inversion techniques for $\delta$ Scuti stars, 
which pose similar observational problems.

\subsection{Rotation in white dwarf stars}

We have much to learn about the global properties, let alone the detailed
interior properties of white dwarfs.  Surface rotation rates are known for a
very small number of white dwarfs.  Rotational broadening in even rapidly
rotating white dwarfs is difficult to measure; only in a few cases is there a
sharp NLTE core in the hydrogen lines in which the effects of rotation may be
detected (Pilachowski \& Milkey 1984,  Koester et al. 1998).  Magnetic white
dwarfs occasionally show rotational modulation (Schmidt 1995).  Fortunately,
rotation is relatively easy to measure in the pulsating white dwarfs; indeed,
the most accurate measures of white dwarf rotation have come from pulsation.
Furthermore, a simple demonstration that pulsating white dwarfs do not rotate
as solid bodies in their interiors would be the first such direct
demonstration for stars other than the Sun.  Even a coarse measure of
differential rotation can provide important constraints on the prior
evolution of a white dwarf, and on conditions in the interior of its
progenitor.

In this paper, we examine the prospects of measuring differential rotation in
white dwarf stars using asteroseismic data.  We find that we can indeed place
important constraints on the internal rotation of stars such as PG~1159,
PG~2131, and GD~358.  With anticipated improved accuracy in the measured
rotational splitting of $g$-mode frequencies, we may indeed see the clear
signature of differential rotation.  In the next section, we explore the
expected splittings for a variety of rotation laws through forward
calculations.  It is important to illustrate a variety of examples, in order
to provide some sense of the sensitivity of the frequency splitting to
$\Omega$, and hence of the diagnostic power of the splitting data.  The
subsequent section reviews the methods for rotational inversion that have
been used with some success in helioseismology, and which can be adapted to
the study of white dwarf stars.  Section 4 shows examples of how these
methods work for models of PG~1159 stars and DBV stars with the synthetic
rotational splittings from Section 2.  In Section 5 we apply these methods to
the published splittings for PG~1159 and GD~358 stars.  We illustrate the
difficulties that are encountered in using the splittings to constrain the
actual rotation rates, and we suggest a model-independent technique for
estimating ${\rm d}\Omega/{\rm d}r$.  We conclude in Section 6 with prospects
for the future progress.

\section{Rotational splittings of oscillation frequencies for various
rotation profiles: the forward problem}

In the forward problem, equation (5) is used to compute the expected
splittings for a given value of $\Omega(r)$.  In practice, the integral is
replaced by an appropriate finite-difference representation, and the kernels
and the angular velocity are evaluated over a grid in $r$. In pre--white
dwarfs,  as in the Sun, radius provides a good independent variable, as the
rotation kernels are evenly sampled throughout the star for modes of
interest.  In the pulsating DB models, radius evenly samples the kernels in
the outer layers, where their amplitudes are largest (see figures 1 and 2).

If there are $M$
independent splittings measured, then equation (5) becomes
\begin{equation}
\delta \nu_i = \sum_{j=1}^{N} w_j K_{ij} \Omega_j \ \ \ \ i=1...M~,
\end{equation}
where the index $j$ corresponds to the radial mesh point in the model,
$w_j$ is a weight which depends on the difference scheme adopted,
and the index $i$ labels the specific splitting.  

In this section, we use equation (6) to explore how $\delta \nu_i$ depends on
various rotation laws, using kernels computed from models of PG~1159 and
GD~358.  This provides some insight into the sort of signal that is to be
expected from white dwarf stars.  We shall then use these synthetic
splittings in the subsequent sections to test various inversion procedures.

Figure 1 shows representative rotation kernels computed from a model by
Kawaler \& Bradley (1994) of the hot white dwarf PG~1159.  The selection of
overtones spans the pulsation modes observed in this star by WET.  Note that
the rotation kernels are remarkably similar, and that they have significant
amplitude over the entire stellar interior.  As discussed in Kawaler et al.
(1985), stars like PG~1159 are indeed pre-white-dwarf stars; while their
interiors are degenerate, the Brunt-V\"ais\"al\"a frequency is not
insubstantial, and the pulsation amplitude in the interior, while smaller
than that at the surface, is much larger than it is in cool white dwarf
stars.  The similarity amongst these kernels renders localization extremely
difficult, if not impossible, and severely limits the power of inversion
techniques to resolve the rotation profile.
\placefigure{fig01}

Figure 2 shows rotation kernels for a model of the pulsating DB white dwarf
star GD 358, plotted in just the outer 20 per cent by radius of the star.  In
this model, the eigenfunctions are relatively much smaller in the stellar
interior: the modes corresponding to the observed frequencies sense the
rotation mainly in the outer layers of the star.  Note that the kernel
amplitude changes dramatically from mode to mode; this is a manifestation of
mode trapping.  The composition transition between pure helium near the
surface and the C,O-rich core (which extends to a fractional radius of 0.98)
is sufficient to pinch those eigenfunctions that have small vertical
displacements at the transition.  These modes have substantially smaller
amplitudes below the transition, and their kernels therefore have much larger
surface values.  From inspection of Figure 2, it is apparent that the trapped
modes have $n=9,10$ and $n=15,16$.
\placefigure{fig02}

Mode trapping is also present in the PG~1159 model, caused by a composition
transition at $r/R=0.72$.  In the kernals shown in Figure 1, this results in
a small bump in the kernels for $n=20$, 24, 28 and 32 at that radial
position -- these are the trapped modes.   For PG~1159 stars, the influence
of mode trapping on the pulsation spectrum, and in particular on the spacing
in period of $g$-modes, is discussed by Kawaler \& Bradley (1994 and see
below).

As long as $\Omega$ changes smoothly,  the rotational splittings that result
depend principally on the envelopes of the kernels.  Since these are quite
similar for all modes in Figure 1, and all modes in Figure 2, save for the
relatively minor evidence for trapping, detailed inversion is bound to be
difficult.

\subsection{PG~1159 stars}

As a first example, we consider the uniform rotation profile with
$\Omega/2\pi = 10$ \uhz.  The spacing of the triplet dipole components are
shown in the top panel of Figure 3.  Note that the splittings are not
constant; this is a measure of the departure of the splittings from the
approximation given by equation (2), and results from the influence of the
vertical displacement on the phase retardation produced by the Coriolis force
in the rotating frame.  The values do indeed approach the asymptotic limit
(2) as $n$ increases.
\placefigure{fig03}

The middle panel in Figure 3 shows the splittings (joined by continuous
lines) that result from a rotation rate that decreases linearly with radius
from 20 \uhz at the center to 0 \uhz at the surface; note that although this
curve has a mean value of a little more than 5 \uhz, as does the uppermost
panel, the splittings vary substantially about that mean.  The dashed lines
in this panel join splittings for another rotation profile that decreases
linearly with radius, but with a shallower slope (from 11.5 \uhz at the
center to 8.5 \uhz at the surface).  In this case, the variations have much
smaller amplitude, but they have exactly the same phase as those produced by
the steeper profile.  The bottom panel is similar to the middle panel, but
shows the splittings for two cases in which the rotation rate increases
linearly with radius.  Notice that the sign of the oscillations about the
smoother dashed curve is reversed.

The periodic variations of the rotational splitting are the consequence of
mode trapping.  Subsurface composition transitions can trap some nonradial
modes between the transition zone and the surface.  The composition gradient
leads to a density gradient, which acts as a reflecting boundary.  Modes with
radial displacement eigenfunctions that are close to zero near the transition
are pinched at that point, and have nearly zero amplitude below.  Because
such trapped modes are confined in the outer nondegenerate layers, they have
substantially smaller kinetic energies at given surface amplitude than do
similar nontrapped modes.  Mode trapping manifests itself in reduced period
spacings; in general, the trapping pushes trapped modes to longer periods
(Kawaler \& Bradley 1994).  The observational diagnostic of trapped modes is
the occurrence of local minima in the period spacing considered as a function
of period.  Figure 4 shows that spacing as a dashed line; the rotational
splittings for the case of a linear decreasing rotation rate are shown by
filled circles connected by continuous lines.  Clearly, the rotational
splittings show the signature of mode trapping.  Since trapped modes are more
confined to the surface, they are more sensitive to rotation in the outer
layers.  For a radially decreasing rotation profile, the trapped modes
exhibit minima in the rotational splitting.  \placefigure{fig04}

Another forward calculation of rotational splitting for the PG~1159 model is
shown in Figure 5.  In this calculation, we computed the rotational splitting
for two different forms of a decreasing rotation law: one that decreases
linearly with radius, and another that decreases as a power--law in $r/R$.
The splittings exhibit the characteristic structure of a rotation rate that
decreases outward.  Despite the rather different rotation laws, the general
forms of $\delta \nu$ as a function of period for both rotation law are
strikingly similar.  The values of the parameters of these rotation curves
were chosen to match the mean value of the rotational splittings of the two
cases.  The maximum difference in splitting is only about 0.03 \uhz.  The
reason for their similarity is simply that the oscillatory behavior of the
splittings is a product principally of the rotationally unperturbed
eigenfunctions, and not of the detailed form of $\Omega$.  This illustrates
that the influence of mode trapping is similar for these two rotation curves,
both of which decrease with radius.  Thus the observed phase between the
rotational splitting and the period spacings provides an observational
determination of the sign of the slope of the rotation curve.  While the
splittings can indicate that the rotation rate decreases outward, we expect
that the observed values will, for PG~1159 stars, be inadequate for detailed
constraints on the form of the rotation law.
\placefigure{fig05}

\subsection{Pulsating DB white dwarfs}

We used the same rotation laws in the models of GD 358 to see the 
response of the rotational splittings.  In this case the response to a 
flat rotation curve is very similar to the case for PG~1159 models 
above; the splittings are shown as the middle (solid) line in Figure 6 
for a flat rotation curve with $\Omega$ of 10 \uhz.  The splittings 
for $\Omega$ decreasing linearly from 11.5 \uhz at the center to 8.5 
\uhz at the surface are shown in Figure 6 by the bottom (solid) line; 
the top (dashed) line shows splittings for a rotation rate that 
increases linearly from 8.5 \uhz at the center to 11.5 \uhz at the 
surface.
\placefigure{fig06}

A glance at Figure 2 shows that the kernels in the GD 358 models sample the
outer layers much more strongly than the core, unlike the PG~1159 kernels
which sample the whole star.  Thus a linear decrease in $\Omega(r)$ results
in an overall shift downwards for the splittings, as the surface layers in
this case are all at a lower value of $\Omega$ when the mean value of
$\Omega$ remains unchanged.  Similarly, splittings computed for a rotation
law that increases with increasing $r$ are displaced upwards.  Beyond this
overall displacement, however, this simple form for the rotation law results
in only subtle changes in the value of $\delta \nu$ as a function of period.
Linear rotation laws change by only a few per cent across the region where
the kernels are large.  Nevertheless, the slight differences
between the splittings of the varying and constant rotation profiles
reverse as the gradient of $\Omega$ is reversed, as for the PG~1159
stars.

Figure 7 shows the rotational splittings for a rotation rate that is the same
as shown in Figure 5 for the PG~1159 model; that is; flat near the center and
then decreasing as $r/R$ to the 10th power to reach a surface value 2 \uhz
smaller.  The central rate was adjusted again to give splittings nearly the
same as for the case of a linearly decreasing rotation curve.  As in the case
for the PG~1159 model, the rotational splittings are very similar; they show
the same pattern for decreasing rotation curves, and differ by small amounts
only.  Thus for GD~358 stars too, it may also be difficult to attain the
necessary accuracy for detailed study of the rotation rate with depth unless
very high accuracy frequencies are available.  As we see below, the results
of inversions are very sensitive to errors in the measured splittings, as is
clear from the results of these forward calculations.  
\placefigure{fig07}

\section{Rotational inversion techniques}

In exploring inversion procedures, we borrow heavily from the field of
helioseismology.  An extensive literature on the inverse problem for solar
rotation exists; as an example, see Gough (1985) and Christensen-Dalsgaard et
al.  (1990).  In this section we assess several procedures, in use in studies
of the solar internal rotation, for possible adaptation to
leuconanoseismology.  In the process, we illustrate how the restrictions
imposed by the observational limitations make very different demands on the
numerical procedures from those in the solar case.  We conclude this section
by suggesting a technique for using inversion theory to best advantage for
white dwarfs.

\subsection{Direct inversion}

Direct solution of the integral equation (5), or its discrete analog equation
(6) (one knows the splittings from observation, and the kernels from theory)
can in principle yield $\Omega(r)$.  The system of $M$ equations contains $N$
unknowns (the values of $\Omega_j$). Hence, unless the number of measured
splittings is greater than the number of resolution points of $\Omega(r)$,
the system of equation is (formally) underdetermined.

In such situations (particularly where $M\ll N$) $\Omega(r)$ need further
constraint.  As described earlier, additional information must be provided.
Alternatively the function space may be restricted by imposing, for example,
a degree of smoothness (see the discussion of regularization below).  In the
opposite case, $M > N$ (such as is frequently the case for the Sun), we
cannot guarantee to satisfy all the splitting constraints.  Under this
circumstance we appeal to the $\chi^2$ statistic, and the function
$\Omega(r)$ that minimizes $\chi^2$ for the inverted splittings is sought
(this is what is usually called least-squares fitting).  Forming $\chi^2$ as
the sum of squares of the differences between the measured splittings and
those computed with the putative solution $\Omega(r)$, weighted with the
inverse square of the standard errors $\sigma_i$ (we presume here that the
errors in the measurement are independent), the minimum in $\chi^2$ is
attained when
\begin{equation} \sum_i \frac{1}{\sigma_i^2} \delta \nu_i w_j
K_{ij} = \sum_i \frac{1}{\sigma_i^2} w_j K_{ij} \sum_k w_k K_{ik} \Omega_k
\ \ \ j=1...N~.
\end{equation} In matrix form, this equation is
\begin{equation}
{\bf K}^T \frac{{\bf \delta \nu}}{\sigma} = {\bf K}^{\rm T}
{\bf K} {\bf \Omega}~,
\end{equation}
where $({\bf K})_{ij} = w_jK_{ij}/\sigma_i$.  This is an $N \times N$ system,
the solution of which provides the values of $\Omega_i$ that minimize the
error-weighted residuals between the observed splittings and those obtained
using $\Omega(r)$ in the forward problem.

As it stands, the equation above is difficult to solve.  If $M < N$ the matrix
${\bf K}^{\rm T} {\bf K}$ is singular.  And even if $M>N$,  the matrix is
very nearly singular because the envelopes of the kernels are similar.  This
singular nature is not particular to white dwarfs, but is inherent in most
inversions in general.

\subsection{Direct inversion with regularization}

To stabilize the solution obtained by least-squares fitting, we must
impose additional conditions on the solution.  If this could be accomplished
in a physically sensible way the outcome could be a rotation curve that is
realistic.  But without a reliable dynamical theory, such imposition is
out of reach.

Instead we resort to regularization, which involves adding
extra terms to the right hand side of equation (8) to penalize
wild behavior of $\Omega(r)$:
\begin{equation}
{\bf K}^T \frac{{\bf \delta \nu}}{\sigma} = 
          ({\bf K}^{\rm T} {\bf K}+ \lambda{\bf H}){\bf \Omega}~.
\end{equation}
The form of the $N \times N$ regularization matrix ${\bf H}$ 
determines what is penalized.  For example, to penalizes steep 
gradient in $\Omega(r)$, ${\bf H}$ is chosen in such a way that ${\bf 
\Omega}^T{\bf H}{\bf \Omega}$ represents, e.g., the squared integral 
of ${\rm d}\Omega/{\rm d}r$ in a finite difference scheme 
(first-derivative smoothing).  In this case it is found that ${\bf 
H}={\bf D}^{\rm T} {\bf D}$, where the matrix $D$ is a discretized 
differential operator (Hansen, Sekii, \& Shibahashi 1992).  In other 
words, the regularization matrix ${\bf H}$ contains terms along (and 
parallel to) the diagonal.  (To obtain approximate first-deriveative 
smoothing, for example, the matrix $D$ is everywhere zero except for a 
line of elements of value $h_j$ immediately above the principal 
diagonal and another line of value $-h_j$ immediately below the 
principal diagonal, where $h_j=1/\sqrt{2(r_{j+1}-r_j)}$.) The 
regularization parameter $\lambda$ is tunable; setting it to zero 
recovers the unstable least-squares formulation.  In practice, 
numerical experimentation with different values of $\lambda$ provides 
the inverter with the information to choose an optimal value, which 
provides a compromise between undue susceptibility to data errors and 
excessive smoothness.

The regularized least-squares fitting technique can be valuable, provided
that it is appreciated how the regularization artificially constraint the
solution.  However, the results can be dreadfully unphysical, especially in
the presence of noise --- in the raw form of the procedure, no a priori
limitations on the value of $\Omega(r)$ are made outside of the
regularization, and with inadequate data it may provide the dominant
influence on the form of the solution.

\subsection{Inversion onto a fixed functional basis: ``function fitting''}

The limited information currently available from oscillation of white dwarf
stars is adequate to determine only a very few properties of $\Omega(r)$.  As
we show in Section 4, attempts at regularized least-squares inversions with
realistic simulations can yield markedly unrealistic results if the actual
rotation rate does not conform to the regularization. Making explicit
assumptions about the form of the rotation law (e.g. that it be positive
definite, and that it either increases or decreases monotonically) we can at
least restrict the class of acceptable functions in an easily comprehensible
way.  We term this ``function fitting.''

One way in which this can be accomplished is to consider projections
of $\Omega(r)$ onto a fixed set of basis functions.  Assume that the angular
velocity in a white dwarf follows the general functional form
\begin{equation}
\Omega(r_j) = \sum_{k=1}^{K} a_k F_k(r_j)~,
\end{equation}
where the $F_k$ are a set of $K$ functions, evaluated of the $N$ radii $r_j$,
and $a_k$ are coefficients to be evaluated.  For example, if the $\Omega(r)$
were assumed to be a simple polynomial in $r$, the functions $F_k$ would be
$r^{k-1}$.  The splitting for multiplet $i$ is then
\begin{equation}
\delta \nu_i = \sum_{j,k}K_{ij}w_{j}F_{jk}a_k~,
\end{equation}
where $F_{jk}=F_k(r_j)$, from which we form a new matrix
$A_{ik}=\sum_j \sigma_i^{-1}K_{ij}w_{j} F_{jk}$
which operates directly on the coefficients $a_k$. Then
the system becomes, in matrix notation,
\begin{equation}
{\bf A}^{T} \frac{{\bf \delta \nu}}{\sigma} = ({\bf A}^{T} {\bf A}) {\bf a}~,
\end{equation}
which is solved for the $k$ elements of ${\bf a}$ in terms of the observed
splittings.  The coefficients ${\bf a}$ then provide that rotation curve, of
the form specified by the equation (10), which fit the data best (in the
least-squares sense).

Of course the inversion discussed in Section (3.1) falls into this category,
with $F_k$ being piecewise constant functions, but since we have in mind
using only a very few such functions we prefer to add the flexibility of
having a more general basis.  We have explored only the very few bases
enumerated below.

\begin{enumerate}

\item simple polynomials of degree $K-1$:
\begin{equation}
\Omega(r)= \sum_{k=1}^{K} a_k (r/R)^{k-1}~;
\end{equation}

\item two linear segments, one with zero slope, joined at $r=r_b$:
\begin{equation}
\Omega(r) = \cases{\Omega_0+\alpha(r_b-r)/R, & $r < r_b$\cr
		   \Omega_0+\beta (r-r_b)/R, & $r \geq r_b$};
\end{equation}
with either $\alpha=0$ or $\beta=0$;

\item constant plus power of radius:
\begin{equation}
\Omega(r)=\Omega_0 + \alpha (r/R)^q
\end{equation}
or
\begin{equation}
\Omega(r)=\Omega_0 + \alpha (1-r/R)^{-q}~,
\end{equation}
where $q$ is positive;

\item piecewise constant functions with a single break:
\begin{equation}
\Omega(r) = \cases{\Omega_1, & $r < r_b$ \cr
		   \Omega_2, & $r \geq r_b$};
\end{equation}

\end{enumerate}

Inversions onto these bases are not linear, for all the bases are 
specified by a parameter that cannot be incorporated into the vector 
${\bf q}$.  In the polynomial and power-law cases, this parameter is 
the degree $K-1$ of the polynomial or the power $q$.  In the forms with 
breaks, the parameter is $r_b$.  To find the best values for these 
parameters, we minimize $\chi^2$ over a range of the nonlinear 
parameter, and selected the solution with the smallest minimum 
$\chi^2$ (see the next section).

\section{Rotational inversions of model splittings}

The results from the forward calculations in Section 2 hinted that 
inversion procedures will be sensitive to the modes measured and the 
accuracy with which the splittings are measured.  In addition, the 
results must depend on the accuracy of the initial equilibrium model 
in comparison with the star itself.  To explore these issues, we now 
calculate inversions using the various inversion procedures described 
above.  We examine how they perform with the synthetic splittings 
computed in Section 2.  We shall discuss inversions of both error-free 
simulated data and simulations that include errors in splittings.

\subsection{Inversion of splittings with no errors}

As discussed in Section 3, direct inversion is an inherently unstable 
calculation; some additional constraints are needed.  To illustrate this 
point, consider the flat rotation curve for the PG~1159 model.  Using all 
of the rotational splittings, a very small value for the regularization 
parameter $\lambda$ of $10^{-4}$, and a regularization matrix corresponding 
to smoothing of the first and second derivatives, the inverted rotation 
profile is shown as a solid line in Figure 8.  The oscillatory behavior of 
$\Omega(r)$ with $r$ is a sign that the numerical instability is not fully 
damped.
\placefigure{fig08}

The dashed line shows the inverted rotation curve with the 
regularization parameter increased to $10^{-2}$; the oscillations are 
considerably damped.  The dotted line shows the results of the 
inversion with a regularization parameter of 1; here the result 
matches nearly precisely to the input rotation curve.  In this figure, 
the departures of the inverted rotation rate from the input rate are 
small; all 21 modes indicated in Figure 3 were used in the 
inversion.  Similar departures between the input and inverted rotation 
profiles were found for the rotation curves that increase and decrease 
linearly with radius.

The flat and linear rotation curves share the property that 
derivatives of $\Omega$ higher than the second are zero; this matches 
the constraints that the regularization matrix places on the solution.  
Because of this it is no surprise that the inversion with 
regularization (and suitable regularization parameter) matches the 
input rotation curve.  We examined splittings cause by the rotation 
curve shown in Figure 5, which is a power law in $r/R$ with a power of $-10$.  
Computation of the inversion of the splittings for this rotation rate 
results in a very precise match with the input rotation curve, again 
using the PG~1159 model.  On a linear scale in radius, the match is 
very good.  Near the surface, the match is still very good, though the 
regularized solution is somewhat lower than the input rotation rate.  
This is because the inversion is enforcing a rotation curve that is 
flat and inflection-free, while the true curve has non-zero 
derivatives at this point.

As a final example of the regularized inversion technique for the 
PG~1159 model, we attempted to invert the splittings in the case of a 
discontinuity in $\Omega(r)$.  The results of this inversion are shown 
in Figure 9 for several different regularization parameters with the 
input rotation rate shown as a heavy line.  In this situation, the 
inversion shows interesting behavior at the position of the 
discontinuity, but only at small values of $\lambda$.  Increasing 
$\lambda$ smoothes the inverted rotation curve, resulting in better 
agreement with the input curve at the center and surface, but washing 
out the signal of the discontinuity at $r/R=0.5$.  In principle, the 
inversion could be better if we adapted the regularization matrix to 
allow for discontinuities.
\placefigure{fig09}

Inversions for the GD~358 model with linear (or flat) rotation curves 
showed the same general features as for the PG~1159 model.  However 
for rotation curves that do not share the properties of the 
regularization matrices, the difference between the input and inverted 
rotation rates was much greater than for the PG~1159 model.  This is 
understandable, given that the rotation kernels for the GD~358 model 
are much larger near the surface; the inversions therefore are 
strongly constrained near the surface, and much less at the center.  
Therefore, for rotation curves such as the power-law rotation curve, 
the inversions will try to fit the surface behavior at the expense of 
the interior.  The response of the inversions to the splittings in the 
case of a discontinuity in $\Omega(r)$ show a similar pattern to that 
illustrated for PG~1159 modes when the discontinuity lies where the 
rotation kernels themselves are large (i.e.  near the surface).  When 
the discontinuity is below this region, the inversion procedure 
matches the flat rotation near the surface, but becomes meaningless 
further in.

\subsection{Inversion of Splittings with Errors}

The experiments recounted above have used synthetic splittings with no
errors.  Of course, observed rotational splittings will contain errors
because of the limited length of observing runs, pattern noise from imperfect
sampling and multiple periods, and noise from photon statistics, seeing, and
other effects.  To examine the effects of observational errors on the
inversions, we have added noise to the computed splittings and calculated
regularized inversions for the PG~1159 model with a flat rotation curve.
Results of this exercise are illustrated in Figure 10.  The noise introduced
was normally distributed (in frequency) with a standard deviation of 0.04
\uhz, which is comparable with the best data available for pulsating white
dwarfs as observed with the Whole Earth Telescope.  Of course, the actual
observed noise is not normally distributed (Costa \& Kepler 1998) but we
choose a normal distribution for illustrative purposes.  The bottom panel of
Figure 10 shows a sample set of splittings with errors along with the
noise-free splittings.  The top panel shows how the regularized inversion
procedure reacts.  Increasing the value of the regularization parameter
$\lambda$ smoothes out the solution, as expected.  With increasing $\lambda$
the inversion approximates the noise free value of a flat rotation rate of 10
\uhz.  However, it is clear that noise introduces false structure in the
inverted rotation curve at the 10\% level, even when the regularization
parameter used is relatively large (0.1).  \placefigure{fig10}

In the case of a non-flat rotation curve, the situation is more 
complicated.  The top panel of Figure 11 shows the results for the 
rotation rate decreasing as the 10th power of $r/R$; the input 
rotation curve is shown as a dark solid line in this figure.  In this 
case, the regularized inversions do detect that the rotation rate 
decreases outward, but do not converge to the proper form.  Increasing 
the regularization parameter smoothes the rotation curve further; it 
approaches to the input rotation curve, but with exaggerated 
structure.  With a regularization parameter of 1, the departure from 
the true rotation curve is smallest, but strong regularization results 
in a rotation curve that drops nearly linearly with radius.  This 
behavior was anticipated given the similarities of the splittings 
shown in Figure 5.
\placefigure{fig11}

We next show how the function fittings behave with noisy splittings.  
The middle panel of Figure 11 illustrates function fittings of the 
same splittings.  Because of the much smaller number of parameters 
being fit, this procedure should be a better choice in this case.  
Since the input rotation curve was one of the four forms used in the 
function fitting, this test may not be fully fair.  However, 
inspection of the figure shows that all forms follow the general trend 
of the input rotation curve, but with exaggerated features.  In this 
sense, the results are similar to the regularized inversion results.  
All show decreasing rotation rates in the outer layers, as was input, 
but all also decrease more than was input.  However, the value of the 
rotation rate near the center is, for all but the polynomial case, 
much closer to the input value.  Of the four forms, the form that 
gives a minimum value for $\chi^2$ was the discontinuous case (the 
thin solid line in Figure 11).

In other cases, the function fitting gives more accurate results.  
Figure 12 show the results of inversions of the splittings (with 
errors) computed using the discontinuous rotation curve in the PG~1159 
model.  The regularized inversions (Figure 12, top) show oscillatory 
behavior for small values of the regularization parameter.  However, 
as $\lambda$ is increased, the inverted rotation curve shows a proper 
decrease with increasing radius, though of course the smoothing 
results in a lack of resolution of the discontinuity.  With function fittings,
the smallest values of $\chi^2$ are obtained for solutions 
using a discontinuous rotation curve; the best curve in the middle 
panel of Figure 12 is such an inversion, and despite the errors in the 
splittings, is a reasonably accurate representation of the input 
rotation curve.
\placefigure{fig12}

In the GD~358 model, the presence of noise also degrades the inversion 
accuracy, as could be expected from examination of Figures 6 and 7.  Figure 
13 illustrates the magnitude of the problem for regularized inversions of 
the flat rotation curve splittings (with noise at the same level as in 
previous figures).  The inverted rotation curves diverge quickly in the 
inner parts for small values of $\lambda$, but even near the surface (where 
the kernel values are largest) the inverted rotation curves are markedly 
different from flat.  Only for $\lambda$ near 1 is the inverted rotation 
curve similar to the input value.  For the function fittings, the results 
also are sensitive to the errors; for this inversion technique, the 
best-fit rotation curve is flat through 98\% of the interior at a rotation 
rate of 10.1 \uhz, and then decreases to about 8.7 \uhz at the surface.
\placefigure{fig13}

\subsection{Errors in the Equilibrium Model}

Another possible source of error is in the background equilibrium 
model used for the inversion: if it is not an accurate representation 
of the star, then the inverted rotation curve will be in error.  To 
illustrate the uncertainties that can arise in the inversion 
procedures from this effect, we recomputed regularized and function fittings
for the flat rotation curve in the PG~1159 model, but using 
a different model for the inversions.  In this test, we used a more 
evolved model (with $T_{\rm eff}=117,100K$, compared to $134,400 K$ 
for the basic model).  This cooler model shows rotation kernels that 
are more concentrated in the outer layers than the original model; the 
periods and trappings are sufficiently different from the first model 
that it would not necessarily be considered an equivalent fit.  This 
model represents a ``poor fit'' to the periods of the main model, but 
allows us to judge the effects of the assumed model on the reliability 
of the results.  Figure 14 shows the results of regularized inversions 
given the error--free splittings from a flat rotation curve.  The 
splittings are from the original model.  Clearly, even with a 
relatively large regularization parameter, the poor model shows 
significant structure in the inverted rotation curve that is entirely 
caused by its improper kernels.
\placefigure{fig14}

Given the results for these tests, it appears that the function 
fitting technique gives reasonably accurate results for the PG~1159 
model in the presence of noise.  However, it is essential to have a 
very good initial equilibrium model that reproduces the observed 
periods and period spacings as precisely as possible.

\section{Application to Observed Splittings in White Dwarfs}

The techniques we have explored show that reliable inversions require a
number of accurately-measured frequency splittings.  The larger the number
of splittings, and the larger the range of $n$, the better.  To date, only
three white dwarf stars have had multiple rotational splittings measured.
These are the GW~Vir stars PG~1159 and PG~2131, and the DB white dwarf
GD~358.  Figure 15 shows the splittings for PG~1159 and GD~358.
\placefigure{fig15}

PG~1159 is the reigning champion, with rotational splittings of 20 $l=1$ 
modes tabulated by WWETPG. Of those 20, 9 multiplets have some uncertainties 
associated with them, either because the multiplets are identified within 
the noise using the spacing constraints or because other mode 
identifications are possible.  The measured splittings are the same to 
within the measurement error.  This suggests that PG~1159 does not have 
significant differential rotation.  Kawaler et al.  (1995) report that 
PG~2131 shows 6 triplet modes, with some uncertainties in splittings 
because of fine structure in the high-frequency component of the triplets.  
The DB white dwarf GD~358 shows 9 triplet modes in the analysis by 
WWETDB.  These show a wide range of splittings, with the splittings 
of low $n$ modes being a factor of 1.6 times smaller than for higher $n$ 
modes.  The simple interpretation, described by WWETDB is 
that GD~358 has a slowly rotating core.

In this section, we apply the tools derived above to the observed spacings in
PG~1159 and GD~358.  As we show, the published splittings in these cases
contain small frequency errors that are fatal to the inversion procedure.
By reanalysis of the original data for PG~1159, however, we use what we have 
learned through the forward problem to show that this star shows a rotation
rate that declines with radius.  We also show that the published splittings 
for GD~358 are suggestive of differential rotation, but that the 
situation is more complex than suggested by WWETDB.

\subsection{PG~1159}

For PG~1159, our starting point is the table of frequencies for 
PG~1159 published in WWETPG.  They 
used the strongest triplets in the power spectrum to determine an 
initial estimate of the period spacing and mean rotational splitting.  
Armed with these, they then dug deeply into the power spectrum to 
identify low amplitude modes.  As a consequence of this analysis, they 
were able to identify probable members of the $l=1$ spectrum; 
however, these low amplitude modes had comparable amplitudes to the 
noise in the spectrum.  Since frequency errors in the splittings can 
cause large errors in the inversions, we cannot include these 
low-amplitude modes in the inversion procedures.  Furthermore, WWETPG 
point out that some of the $l=1$ components may be identified 
incorrectly.

Another problem with the published frequencies is that they were determined
directly from the computed power spectrum.  Therefore, neighboring peaks
(whether from noise or from the star) can ``pollute'' one another, and cause
small errors in the frequency determinations for nearby peaks.  Though the
spectral window was extremely good for this Whole Earth Telescope run,
residual aliases can also affect the accuracy of frequency determinations
when made directly from the power spectrum.  One way to solve this problem is
to do a nonlinear least-squares solution for the amplitude, frequency, and
phase for all peaks in the spectrum simultaneously.  This procedure was used
by Kawaler et al. (1995) for PG~2131, and described in more detail in 
O'Brien et al. (1996).  As long as one is careful not to allow
the nonlinear least-squares procedure to converge on a local minimum in
$\chi^2$, then this procedure allows one to find the frequencies of each peak
as if it were isolated; that is, the effects of sampling and pollution by
other peaks are removed. 

\placetable{tab1}
To illustrate this influence, we show in Table 1 the frequencies for several
modes in PG~1159 obtained from WWETPG, and from a simultaneous nonlinear
least-squares fit to their data.  In the least-squares solution, only the
modes near clearly identified triplets were included; without modes present
that are clearly above the surrounding noise level, least-squares solutions
suffer from the same problem of noise pollution.  The amplitudes in this
table are in milli--modulation amplitude; one mma is $\delta I/I$ of
$10^{-3}$.  Even though the differences in frequency are small (from a
model-fitting perspective) the splittings are significantly different.
Another interesting feature of this comparison is the amplitudes are
significantly reduced in the largest peaks, and that the relative amplitudes
are different.  This is caused by overlapping of the windows of these strong
peaks.  The fact that the amplitude ratios change in the least-squares
analysis is important for interpretation of the amplitudes of the separate $m$
components in terms of inclination of the rotation/pulsation axis.  
Figure 15 shows the splittings computed via least--squares as solid 
dots connected by a solid line.

Using the frequencies in Table 1, we computed the regularized inversions for
PG~1159 using a best-fit model that was very similar to the model of Kawaler
\& Bradley (1994).  This is not the model that was used for the tests in
previous sections, but is optimized to fit the periods and period spacings of
PG~1159 itself.  Figure 16 shows the regularized inversions of the splittings
with $\lambda$ of 0.01, 0.1, and 1.0.  The inversion with the smallest
$\lambda$ shows what looks like spurious structure, based on tests with
sample data.  With a larger $\lambda$, the rotation curve appears to decrease
with increasing radius, with some small contrast.  Is this real?
\placefigure{fig16}

If so, it should also manifest itself in the function fittings.  All of the
forms for the rotation law minimize $\chi^2$ when the surface rotation rate
is smaller than the central value; the best fit of the three forms is the
discontinuous rotation curve, with a surface rotation rate of about 7 \uhz
and a central (within the inner 20\% of the radius) rate of about 12 \uhz.
These numbers are inherently uncertain; such a rotation law is simply
consistent with the the small number of splittings within the errors of the
splittings.

These indications of a decreasing rotation rate from the center to the
surface are model dependent; a slightly different model for PG~1159 that
matches the periods nearly as well could indeed produce a quantitatively
different rotation curve.  However, Figure 3 shows a way in which we can
estimate the sign of the slope of the rotation curve without recourse to
detailed models.  The general features of that figure was that a rotation
curve that decreases with radius shows variations in $\delta \nu$ that are
out of phase with the variations for a rotation curve which increases with
radius.  Mode trapping, as is manifest in departures from uniform period
spacings, provides the key.  Figure 4 shows the variations of $\delta \nu$
for a rotation rate that decreases with increasing radius compared with
departures from uniform period spacing (backward-differenced).  When the
rotation rate decreases with radius, the variations in $\delta \nu$ are
largely in phase with the variations in $\Delta P$.  For an increasing
rotation curve, then, the variations in $\delta \nu$ will be out of phase
with variations in $\Delta P$.  Therefore, for PG~1159 and any other star
that shows a sufficient number of consecutive overtones, plotting the
frequency spacing and period spacings together allows determination of
whether the rotation rate increases or decreases with depth, without recourse
to detailed models.  Such a plot is shown in Figure 17; the decreasing
splittings between 500 seconds and 580 seconds fall in parallel with the
decreasing period spacings.
\placefigure{fig17}

\subsection{GD~358}

GD~358, the prototype of the pulsating DB (helium-atmosphere) white 
dwarfs, has been observed by the Whole Earth Telescope twice.  The 
first WET campaign of May 1990 was described in detail by Winget et 
al.  (1994).  We did not perform a complete least-squares analysis of 
the data from this campaign.  As a result, we will not attempt to 
determine if the splittings are in phase with the period spacings, as 
was possible with the PG~1159 data.  The increase in rotational 
splitting on pulsation period that WWETDB deduced from the power 
spectrum analysis (see Figure 15) will probably not change 
dramatically with a reanalysis.  The two shortest-period modes show 
splittings between 3 and 4 \uhz, while the longer period modes have 
splittings of about 6.4 \uhz.  The results of our experiments with the 
forward analysis (see Figure 7, for example) suggest that producing 
such a large contrast in splittings requires a very steep change in 
the rotation rate in the stellar interior.  The model of GD~358 that 
we use to obtain kernels and attempt inversions is described in Dehner 
\& Kawaler (1995); it represents a good match to the observed periods, 
effective temperature, and luminosity of GD~358.

As a first approximation, we try to invert the observed splittings 
using function fitting.  Contrary to expectation, the best-fit 
rotation curves show a rapidly rotating inner core, with the 
discontinuity at a fractional radius near 0.2.  For a discontinuous 
rotation curve, the best fit is for an inner rotation frequency of 140 
\uhz (out to $r/R=0.20$) with the rotation rate in the remainder of 
the model of 0.6 \uhz.  A similar quality fit was found for a rotation 
rate of 280 \uhz at the center, decreasing linearly to 0.25 \uhz at a 
fractional radius of 0.22 and remaining constant out to the surface.  
A somewhat similar rotation curve is found through a regularized 
inversion (with constraints on the first and second derivatives of 
$\Omega(r)$).

Figure 18 shows the results of a representative solution; the top 
panel shows the rotation curves resulting from a function fitting and 
from the regularized inversion (with $\lambda=10^{-2})$.  The bottom 
panel shows a comparison of the observed splittings with the model 
splittings, computed with a forward calculation, for the rotation 
curves in the top panel.  Clearly, the regularized inversion provides 
an unphysical rotation curve -- a negative rotation velocity through much 
of the interior -- despite producing splittings that match the 
observed splittings quite well.   The feature of the rotation 
curve that causes the low-$n$ modes to have smaller frequency 
splittings is the steep decrease in $\Omega$ in the inner regions.
\placefigure{fig18}

The reason why the inversions result in a rapidly rotating inner 
region result from the combined effects of the behavior of $K$ near 
the center and the effects of mode trapping.  We illustrate these 
effects in Figure 19.  Figure 19(a) shows the running integral 
$\int_{0}^{r} K_{i}(r')dr'$ as a function of $r$ for selected $l=1$ 
modes in our model of GD~358.  The top panel shows the integral over 
the entire model for the trapped mode $n=9$, the untrapped mode 
$n=13$, and the trapped mode $n=15$.  The splitting of the $n=9$ mode 
in GD~358 is 3.6 \uhz, while the splitting of the other two modes are 
6.4 \uhz.  The two bottom panels show the integrated kernels near the 
center and surface.
\placefigure{fig19}

Figure 19(b) shows that the higher-$n$ modes have larger integrated 
kernels than the $n=9$ mode.  Within $r/R=0.2$, the $n=13$ and $n=15$ 
modes have nearly identical values, which are about 40\% larger than 
the value for the $n=9$ mode.  This turns out to be the dominant 
factor that causes the inversion procedures to demand a rapidly 
rotating core.  By isolating fast rotation in these inner regions, the 
inverted rotation curve can produce large splittings for the 
higher-$n$ modes.  But, because these inner regions contribute only 
5\% of the total splitting, the inner rotation rates required are
quite fast (see Figure 18).

The secondary effects of mode trapping are illustrated in Figure 19(c).  
The three low-$n$ modes ($n$=8,9, and 10, with $n=9$ shown in the 
figure) are trapped modes which show a large maximum in the rotation 
kernels near the surface.  Therefore, these modes sense the outer 
regions more than the untrapped modes ($n$=11-13, with $n=13$ shown in 
Figure 19(c)).  If the outer region is rotating much more slowly than 
the inner region, then the trapped modes will show smaller splittings 
than untrapped modes.  For the next set of trapped modes, ($n$=14-16, 
$n=15$ shown in Figure 19(c)), the maximum value of the rotation kernel 
is concentrated farther out than the lower-$n$ trapped modes.  
Therefore, integrating these kernels over the narrower region has a 
smaller impact on the splittings, and the higher-$n$ trapped modes 
show larger splittings than the lower-$n$ trapped modes (beyond the 
effects of the central regions described above.

It is not at all clear how realistic the results of these inversions 
are for GD~358; the numerical procedures attempt to utilize the small 
(in an absolute sense) differences in the kernels near the center to 
minimize $\chi^{2}$ while most of the action in the pulsations is near 
the surface.  If we confine our inversions to adjust $\Omega$ in the 
outer regions only, can the inversions do nearly as well?  Using 
function fittings with discontinuous rotation curves, we show in 
Figure 20(a) the value of $\chi^{2}$ as a function of the fractional 
radius of the discontinuity.  The deepest minimum is at a fractional 
radius of 0.2, and corresponds to the discussion above.  However, 
there is a deep local minimum in $\chi^{2}$ in the outermost layers.  
The corresponding solution has a rotation frequency of 8.4 \uhz from 
the center to a fractional radius of 0.9974; the rotation frequency 
{\em increases} discontinuously to a surface rate of 66 \uhz.
\placefigure{fig20}

Note that this outer solution suggests a fast-rotating surface, while 
the inner solution indicates a fast-rotating core!  Figures 20(b,c) show 
the splittings that result from these solutions, along with the data 
from GD~358.  The near-surface discontinuity does produce splittings 
that are as close to the observed splittings as those from the deep 
discontinuity.  Since the observed splittings that we used in this 
study are taken from the power-spectrum analysis of WWETDB, it is 
still to early to claim with any certainty that GD~358 has either a 
rapidly rotating core or a rapidly rotating outer envelope.  
Additional complications, including the possible effect of magnetic 
fields on the observed splittings in GD~358 (WWETDB, Jones et al.  
1989) further complicate the picture.

\section{Conclusions: Prospects for Future Inversions}

This investigation has shown that even though the information provided 
by the pulsations of white dwarf stars is much less extensive than by 
solar oscillations, some pulsators are rich enough to constrain their 
internal rotation.  The technique of regularized inversion is of 
limited use under even these circumstances, but can provide insights into 
the oscillation properties of models of these stars.  Function fitting,
where the inverted rotation profile is constrained to fit 
simple functional forms, represents a useful alternative.

For the pulsating white dwarf stars, very accurate pulsation 
frequencies are needed to allow constraint of the rotation curve.  We 
explored the data on PG~1159, and were able to demonstrate that the 
rotational splittings are sensitive to mode trapping.  In fact, we 
show that the phase between the rotational splitting and period 
spacings (as functions of period) allows an observational 
determination of the sign of the slope in the rotation rate with 
depth.  PG~1159 shows the signature of a rotation curve which is 
faster at the center than at the surface.  Further observations of 
this star by the Whole Earth Telescope are now available (Winget, 
private communication); least-squares analysis of these data, in 
combination with the first WET data on this star, should be able to 
test this result.

The case of the pulsating DB white dwarf GD~358 is interesting as 
well.  The data available in the literature indicate that this star 
does not rotate as a solid body.  The rotation kernels from Figure 2, 
and the experiments described in this paper, illustrate that earlier
interpretation that the lower-$n$ modes sense the interior 
more than the higher $n$ modes (which is what led WWETDB to conclude 
that the outer layers of GD~358 rotate faster than the inner layers) 
is oversimplified.  As we show, the observed spacings can 
be reproduced if the star has a rapidly rotating inner ($r/R < 0.2$) 
core or, with a bit less accuracy, if it has a rapidly rotating outer 
surface (the outer 0.003 $R_{*}$).  Clearly, we need to reanalyze the 
data from the two WET campaigns on this star with the least-squares 
technique to obtain more accurate pulsation frequencies and splittings.

\acknowledgments

SDK acknowledges the National Science Foundation for support under the 
NSF Young Investigator Program (Grant AST-9257049) to Iowa State 
University.  and the NASA Astrophysics Theory Program through award 
NAG5-4060.  He is also very happy to thank the Institute of Astronomy 
and Churchill College, Cambridge University, for their hospitality and 
support during his visit.  TS is grateful for the support by the 
Particle Physics and Astronomy Research Council.  Mike Montgomery of 
the University of Texas provided several helpful comments on this 
paper, and we also benefited from helpful discussions with Wojtek 
Dziembowski.

\newpage

%\plotone{fig01.ps}
\figcaption[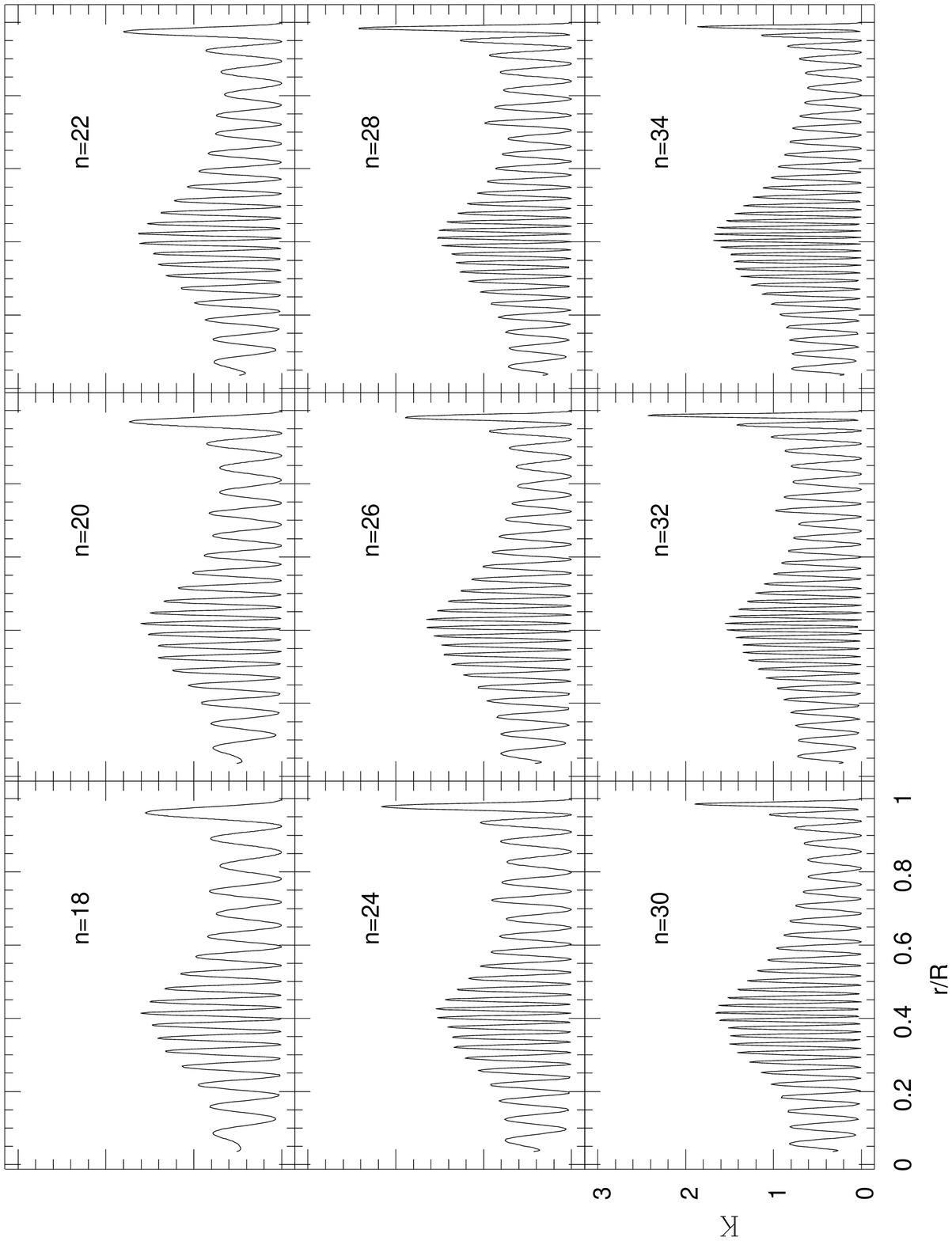]{Sample rotation kernels $K_n(r)$ of $l=1$ $g-$modes
in a model of PG 1159.  As described in the text, the kernels are normalized
by the kinetic energy of the mode. \label{fig01}}

%\plotone{fig02.ps}
\figcaption[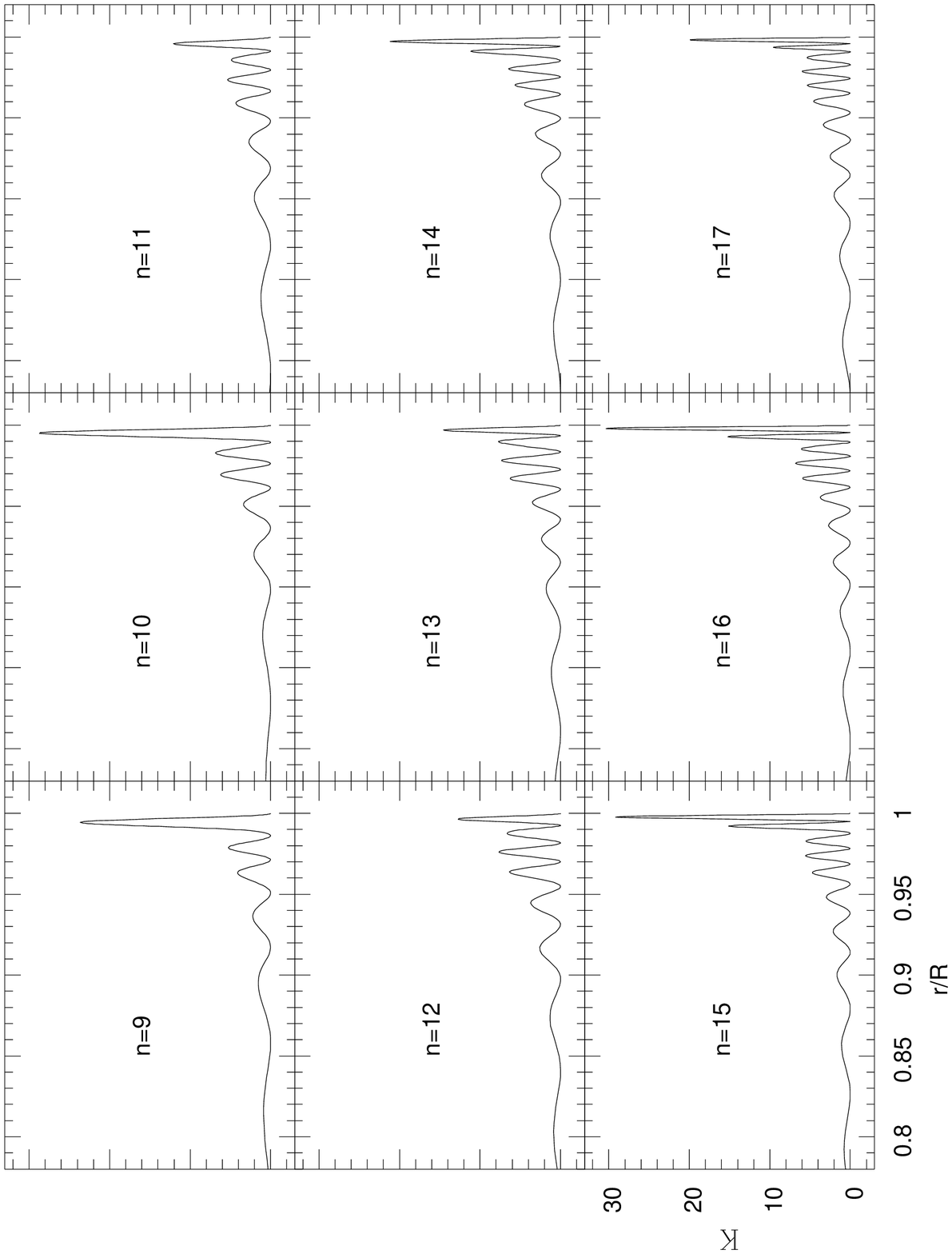]{Sample rotation kernels $K_n(r)$ of $l=1$ $g-$modes
in a model of GD 358. Only the outer 20\% of the radius is shown the
amplitude of the kernel is very small in the interior.  Note that the
amplitude of the kernel in the outer layers is a strong function of $n$; this
is the consequence of mode trapping. \label{fig02}}

%\plotone{fig03.ps}
\figcaption[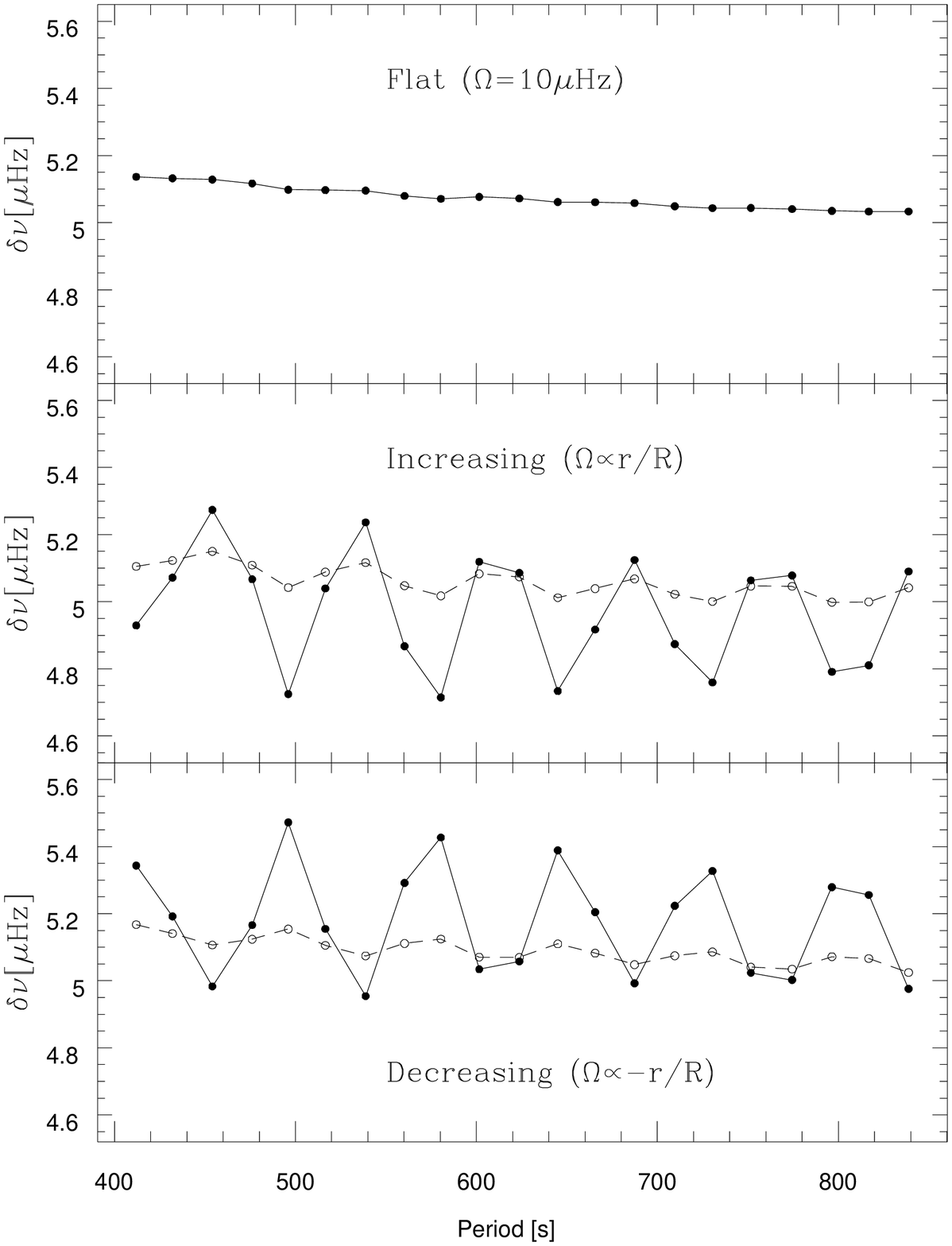]{Rotational splitting (in \uhz) for the PG~1159
model with various forms for $\Omega(r)$.  Top: a flat rotation curve, with
$\Omega=10$\uhz.  Middle: a linearly decreasing rotation curve with a small
slope in $\Omega(r)$ (dashed line) and a steep slope in $\Omega(r)$ (solid
line).  Bottom: a linearly increasing rotation rate with a small slope
(dashed line) and a steep slope (solid line). \label{fig03}}

%\plotone{fig04.ps}
\figcaption[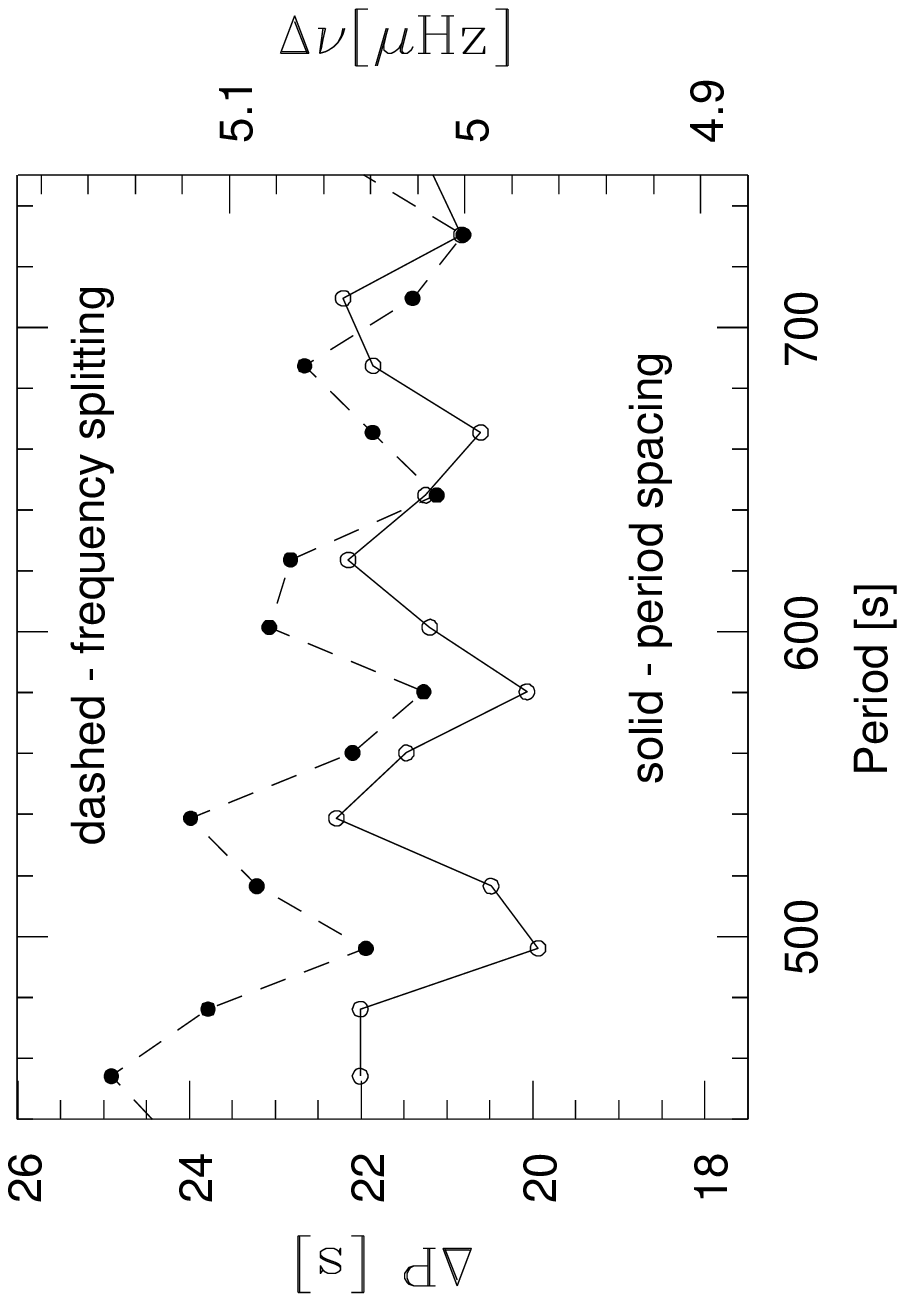]{Period spacing as a function of period in the 
representative PG~1159 model (solid line), with values indicated on 
the left axis.  The dotted line shows the rotational splitting 
computed with a linearly decreasing rotation curve (see Figure 3), 
with values indicated on the right axis. \label{fig04}}

%\plotone{fig05.ps}
\figcaption[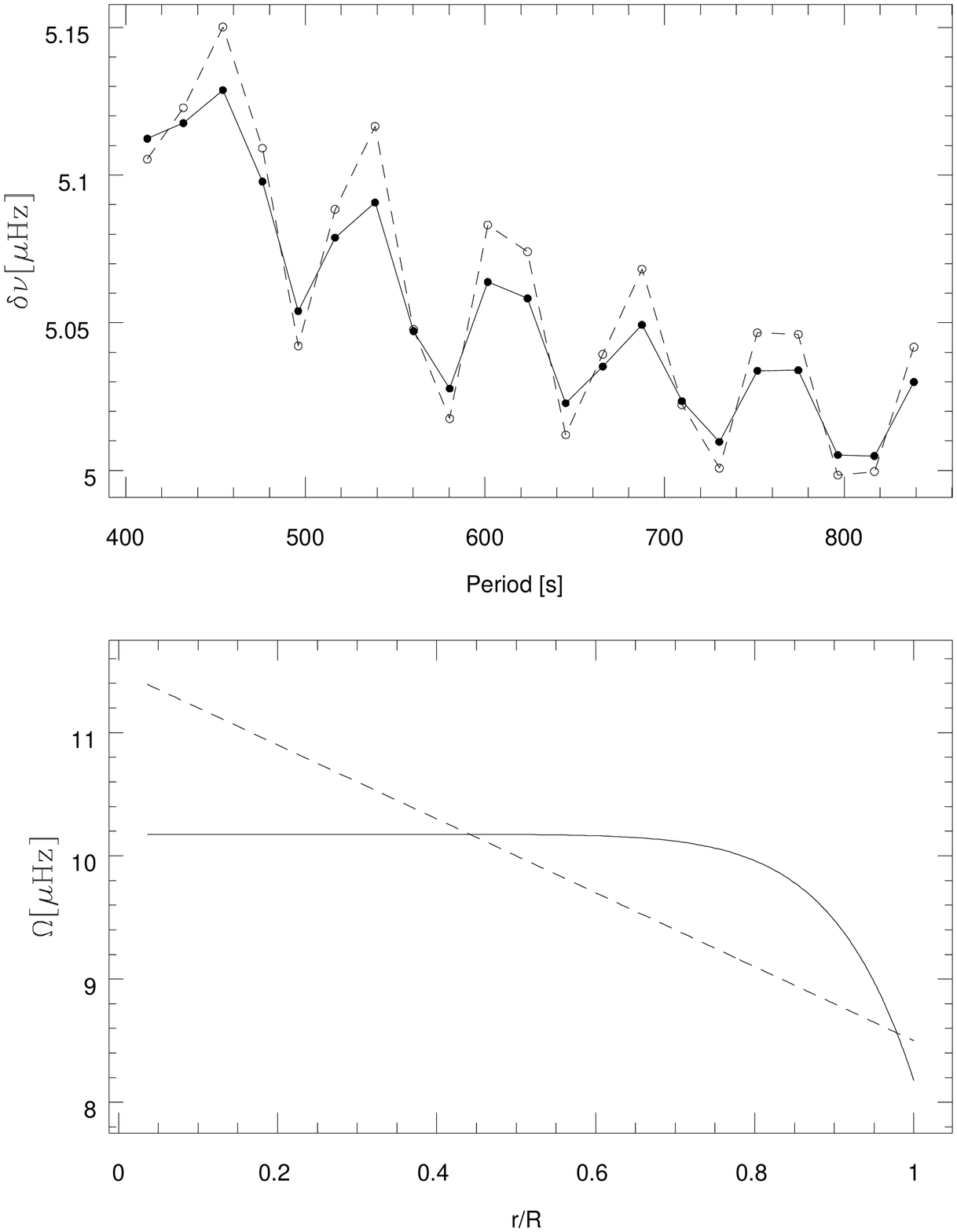]{Top: the dashed line shows rotational splittings
computed for a PG~1159 model with a rotation curve that decreases linearly
with radius (middle panel of Figure 3) compared with the splittings for a
rotation curve that follows a power law (solid line).  Bottom: rotation rates
for the two models. \label{fig05}}

%\plotone{fig06.ps}
\figcaption[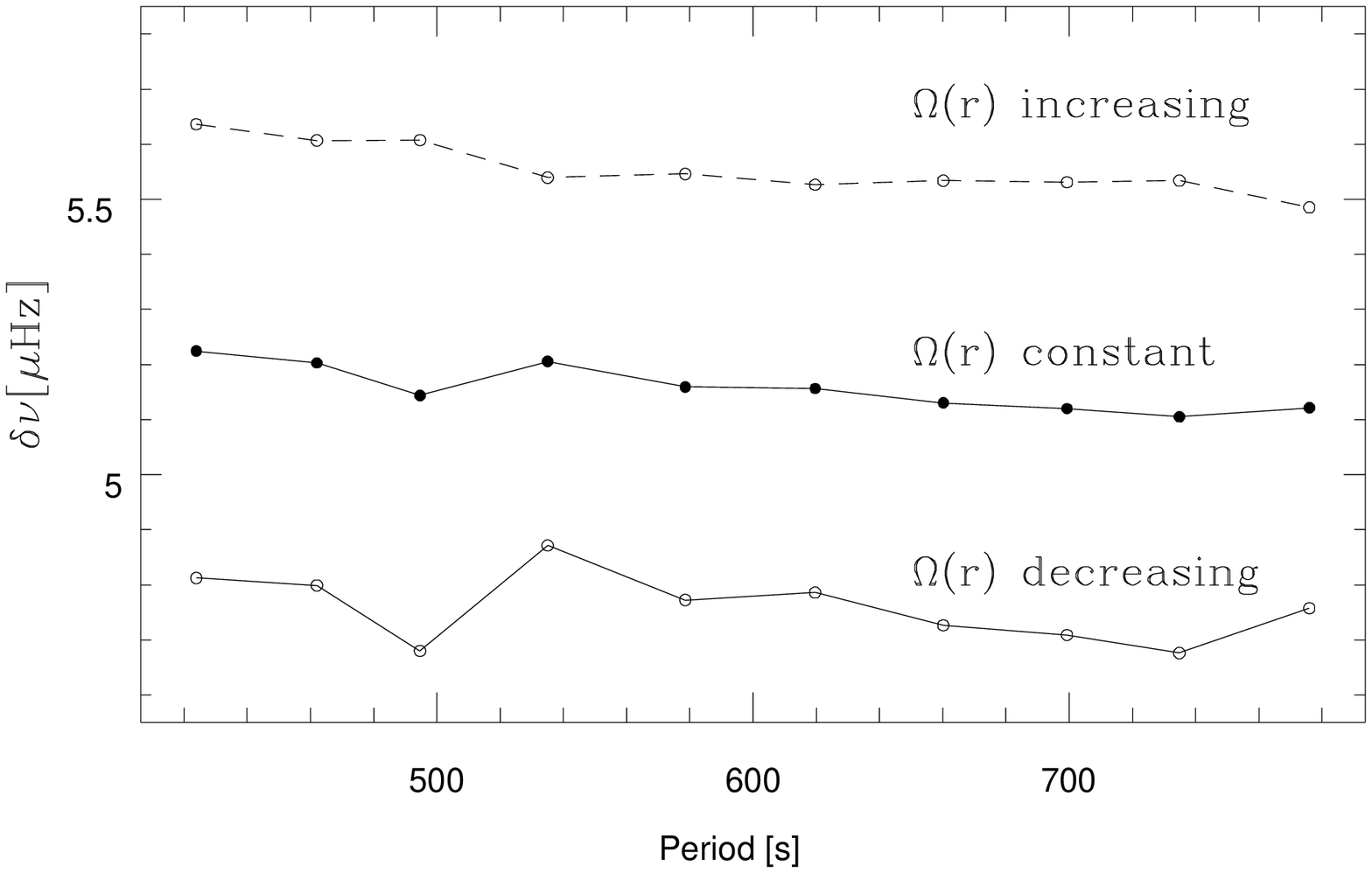]{Rotational splittings for modes in the model of
GD~358.  The solid line show the splittings for a flat rotation curve with a
rotation rate of 10 \uhz.  The top dashed line shows splittings for a
rotation curve that increases linearly with radius, while the lower curve
shows that for a rotation rate that decreases linearly.  All three rotation
curves have the same mean values (mean with radius). \label{fig06}}

%\plotone{fig07.ps}
\figcaption[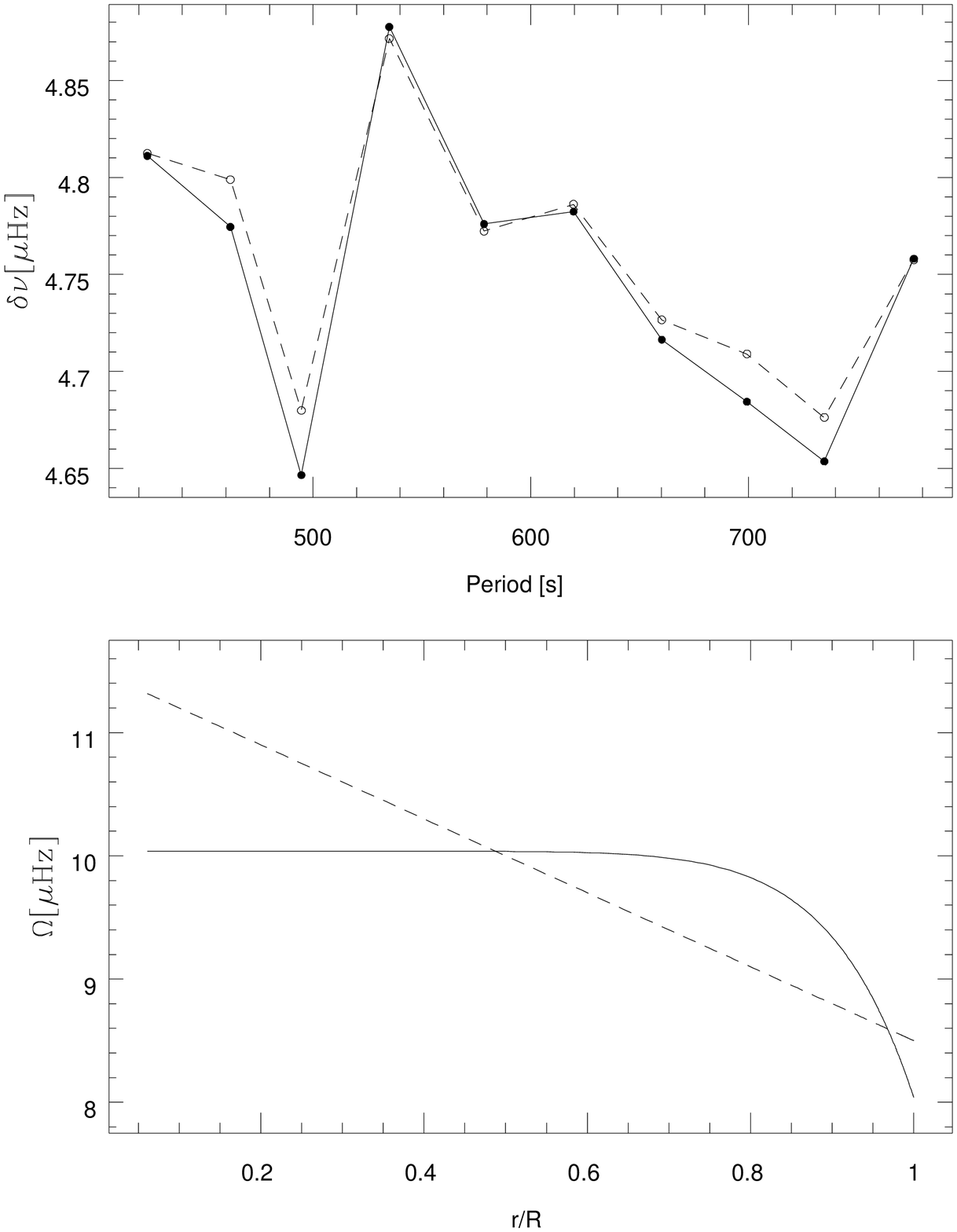]{Top: the dashed line shows rotational splittings
computed for a GD~358 model with a rotation curve that decreases linearly
with radius compared with the splittings for a
rotation curve that follows a power law (solid line).  Bottom: rotation rates
for the two models. \label{fig07}}

%\plotone{fig08.ps}
\figcaption[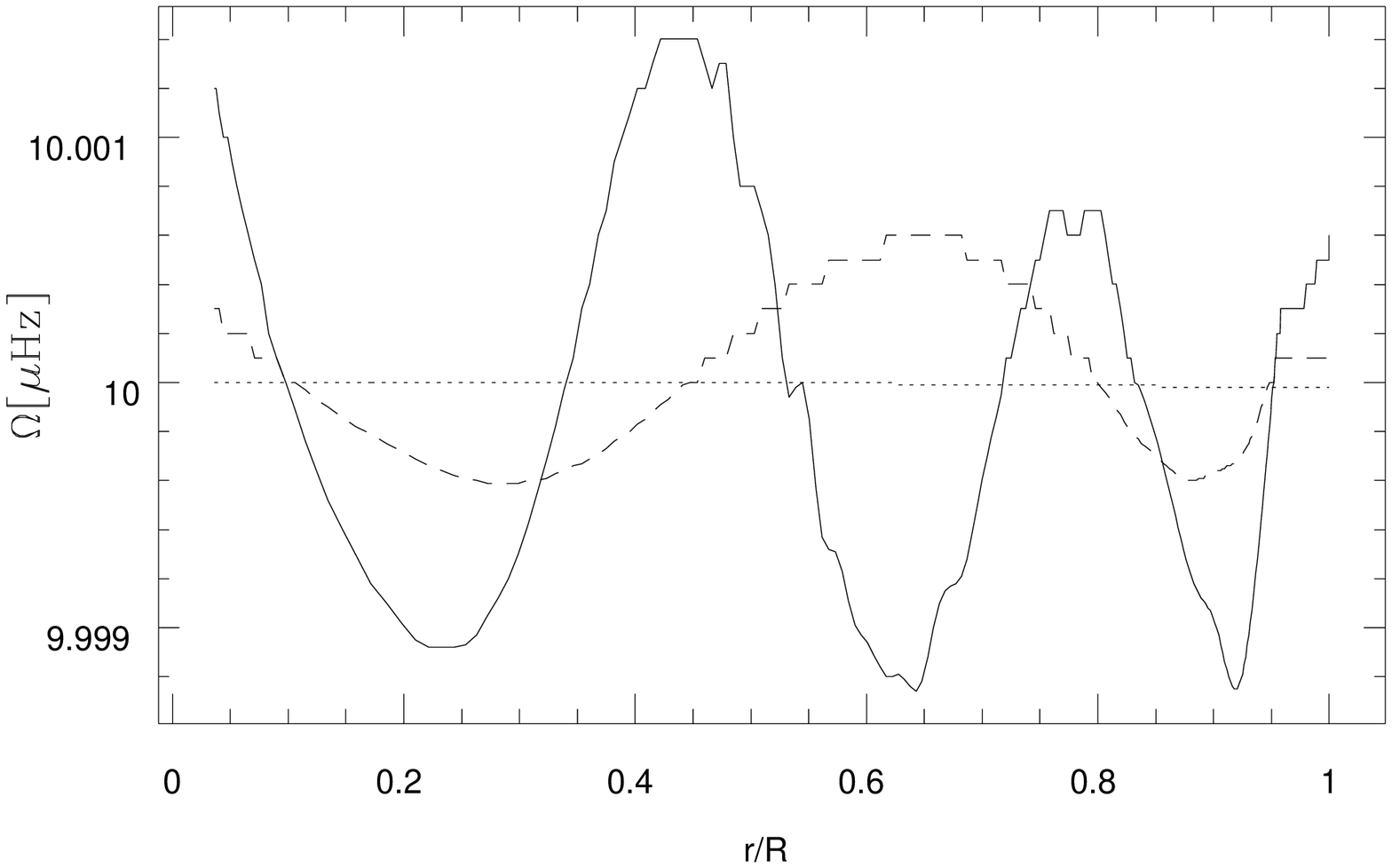]{Results of regularized inversions of rotational
splittings for the PG~1159 model with a flat rotation curve.  The solid line
is for a regularization parameter $\lambda$ of $10^{-4}$; the dashed line
shows the inverted rotation profile for $\lambda=10^{-2}$ and the dotted line
is the inversion with $\lambda=1$. \label{fig08}}

%\plotone{fig09.ps}
\figcaption[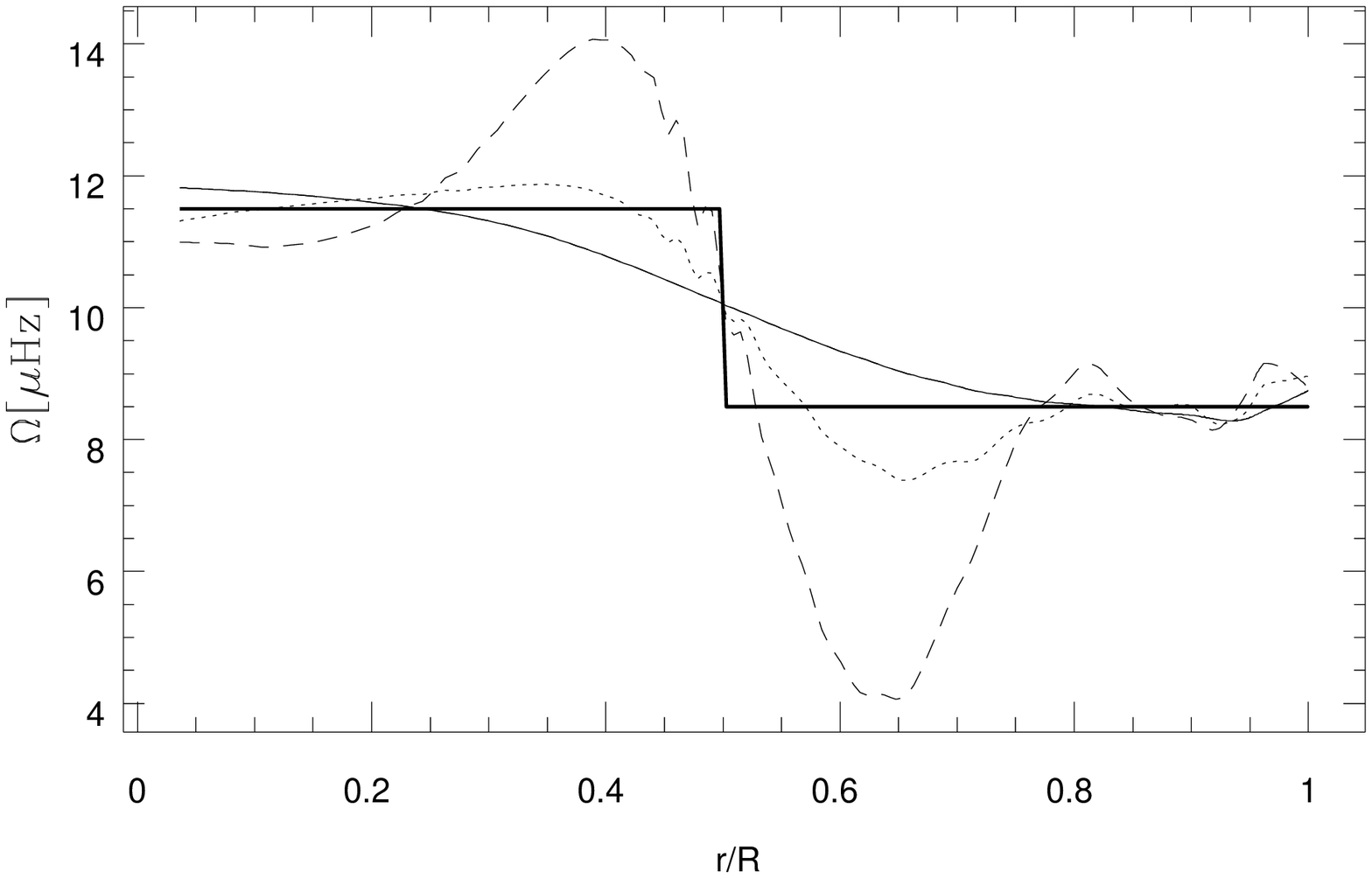]{Regularized inversion for PG~1159 model with a
rotation curve with a discontinuity at the half-radius point (heavy line).
The inversions are with $\lambda=10^{-4}$ (dashed line), $\lambda=10^{-3}$
(solid line) and $\lambda=10^{-2}$ (dotted line). \label{fig09}}

%\plotone{fig10.ps}
\figcaption[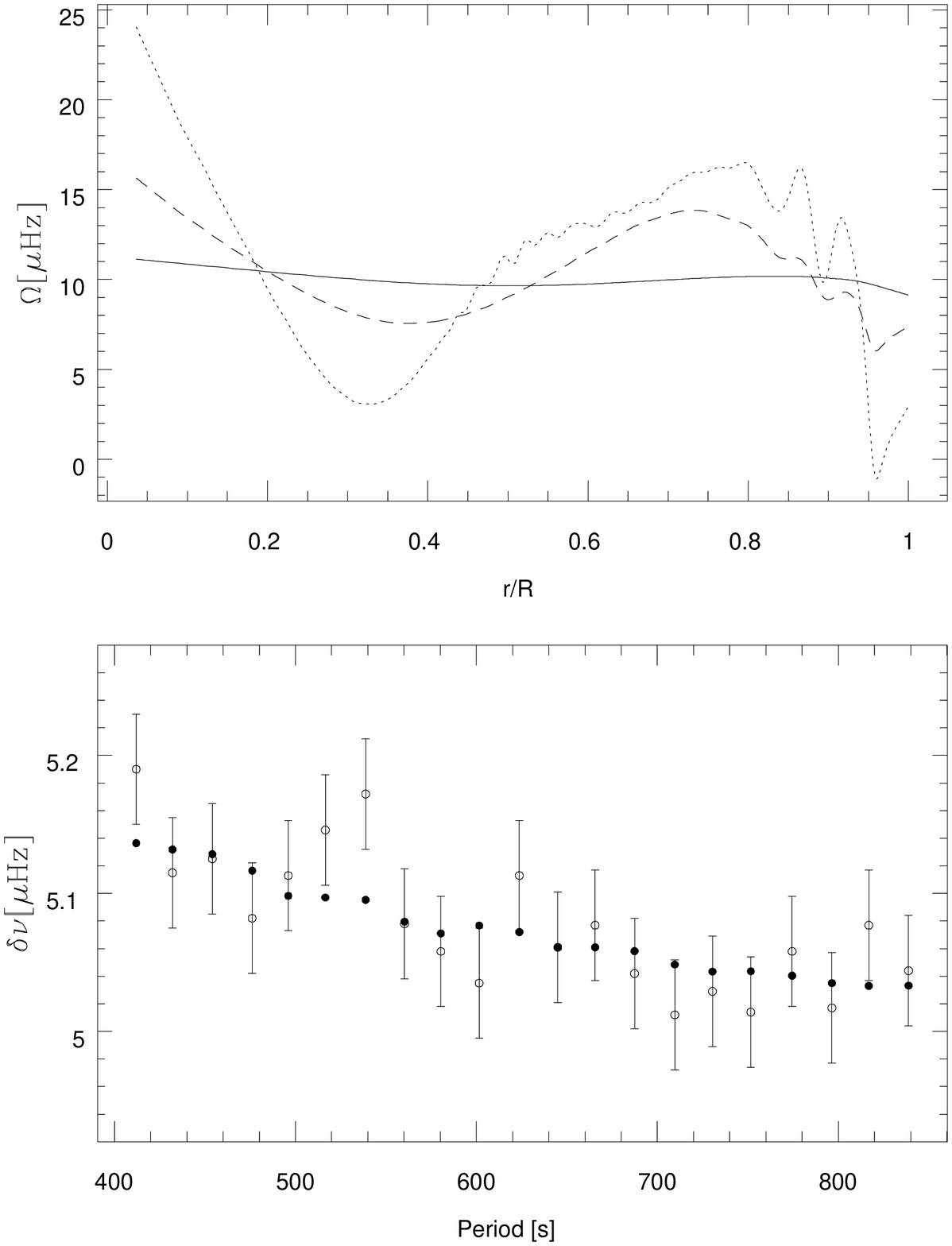]{The top panel shows the inverted rotation curves
for splittings computed with a flat rotation rate, but with noise added to
the splittings.  In this figure, the solid line signals the inversion with a
regularization parameter of 0.1, the dashed line has $\lambda=0.01$, and the
dotted line has $\lambda=0.001$.  The bottom panel shows the splittings for a
flat rotation curve with no noise (solid points) and the same splittings with
a normally distributed error of 0.04 \uhz. \label{fig10}}

%\plotone{fig11.ps}
\figcaption[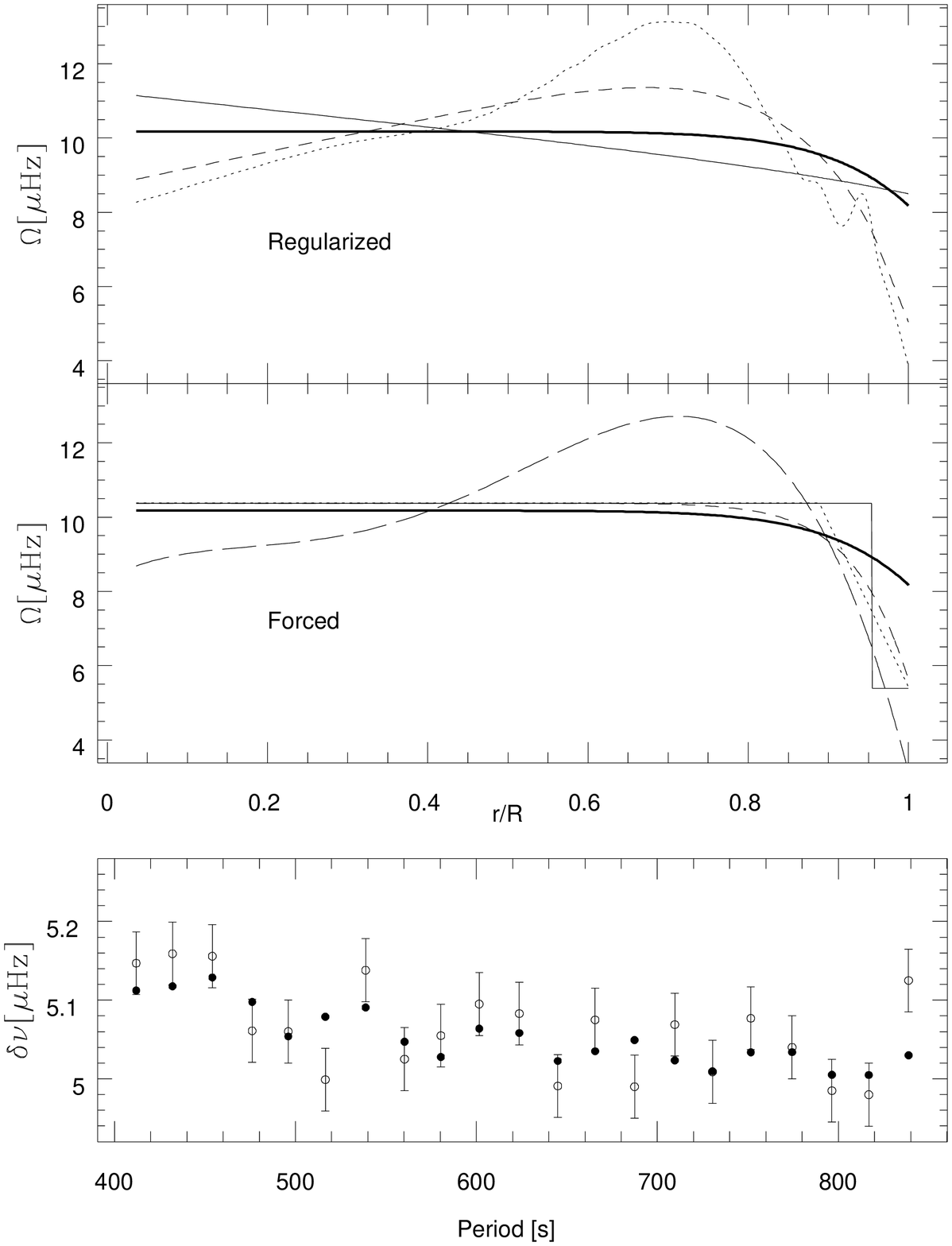]{Inversions of splittings for the rotation curve 
from Figure 5, but including synthetic errors for the splittings with 
a normally distributed error of 0.04 \uhz.  The input rotation curve 
is shown as a dark solid line in the top two panels.  {\em Top:} 
results of regularized inversions of the simulated splittings.  
$\lambda=1$ is the thin solid line; the dashed line has $\lambda=0.1$, 
and the dotted line has $\lambda=0.01$.  {\em Middle:} Function fittings
using the same splittings.  Long-dashed line: polynomial 
rotation curve, degree 5.  Short-dashed line: power-law rotation 
curve.  Dotted line: flat+linear.  Thin solid line: discontinuous 
rotation curve.  {\em Bottom:} error-free splittings as solid dots, 
and the splittings with simulated errors as open circles with error 
bars.  The inversions shown in the top two panels all used splittings 
given by the open circles. \label{fig11}}

%\plotone{fig12.ps}
\figcaption[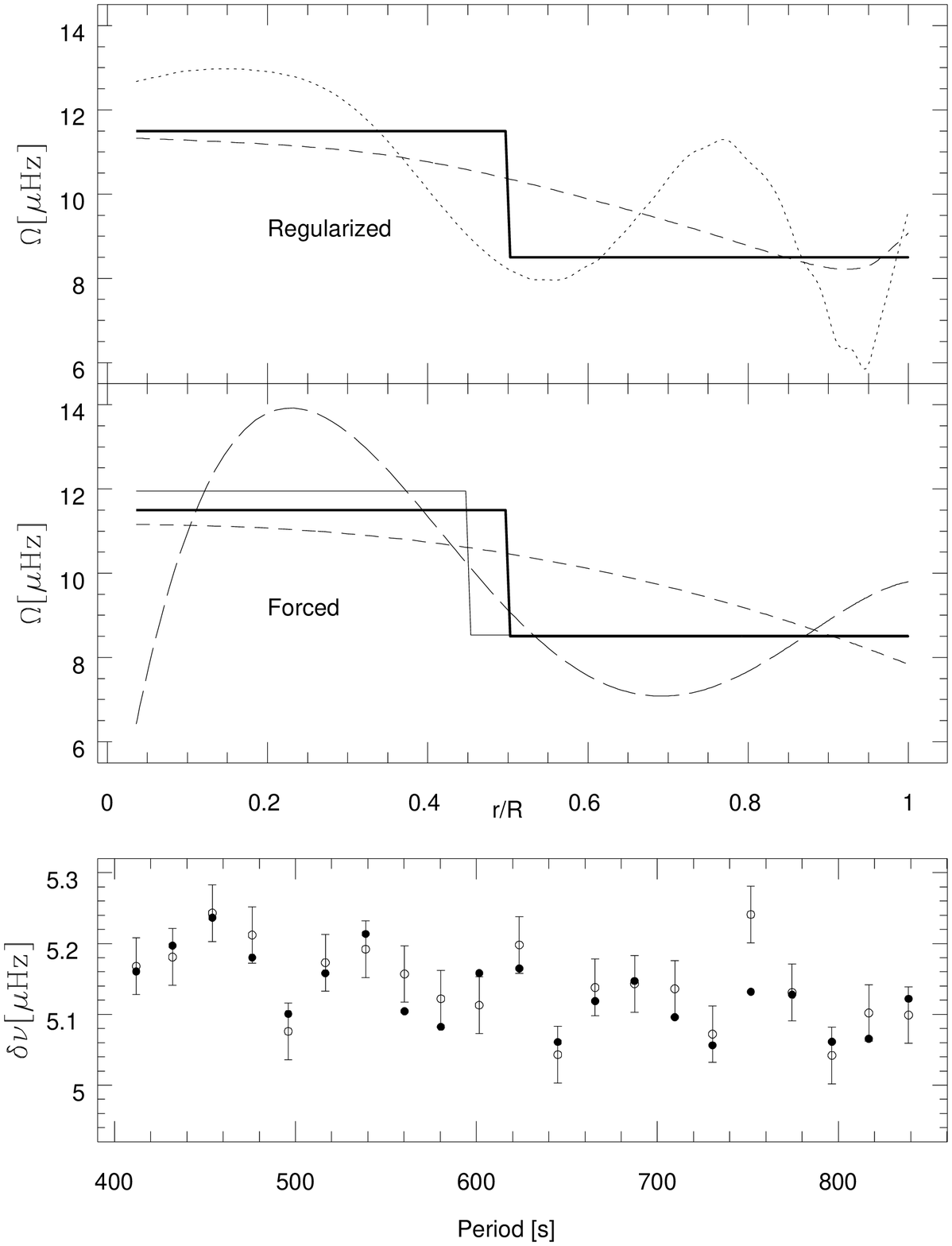]{Inversion of splittings for a discontinuous 
rotation curve in a PG~1159 model.  The input rotation curve is shown 
as a dark solid line in the top two panels.  {\em Top:} regularized 
inversion results with $\lambda=0.01$ (dotted line) and $\lambda=0.1$ 
(dashed line).  {\em Middle:} Function fittings with the same 
splittings.  Long-dashed line: 5th order polynomial.  Short-dashed 
line: power-law rotation curve.  The best value for the power in this 
case is 2.5.  Thin solid line: best-fit rotation curve with a 
discontinuity.  Of the three inversions shown, the discontinuous 
rotation curve had a minimum value of $\chi^2$.  {\em Bottom:} the 
forward-calculation splittings (solid dots), and splittings with 
synthetic errors (normally distributed over 0.04 \uhz).  Inversions 
shown in the top two panels all used splittings given by the open 
circles. \label{fig12}}

%\plotone{fig13.ps}
\figcaption[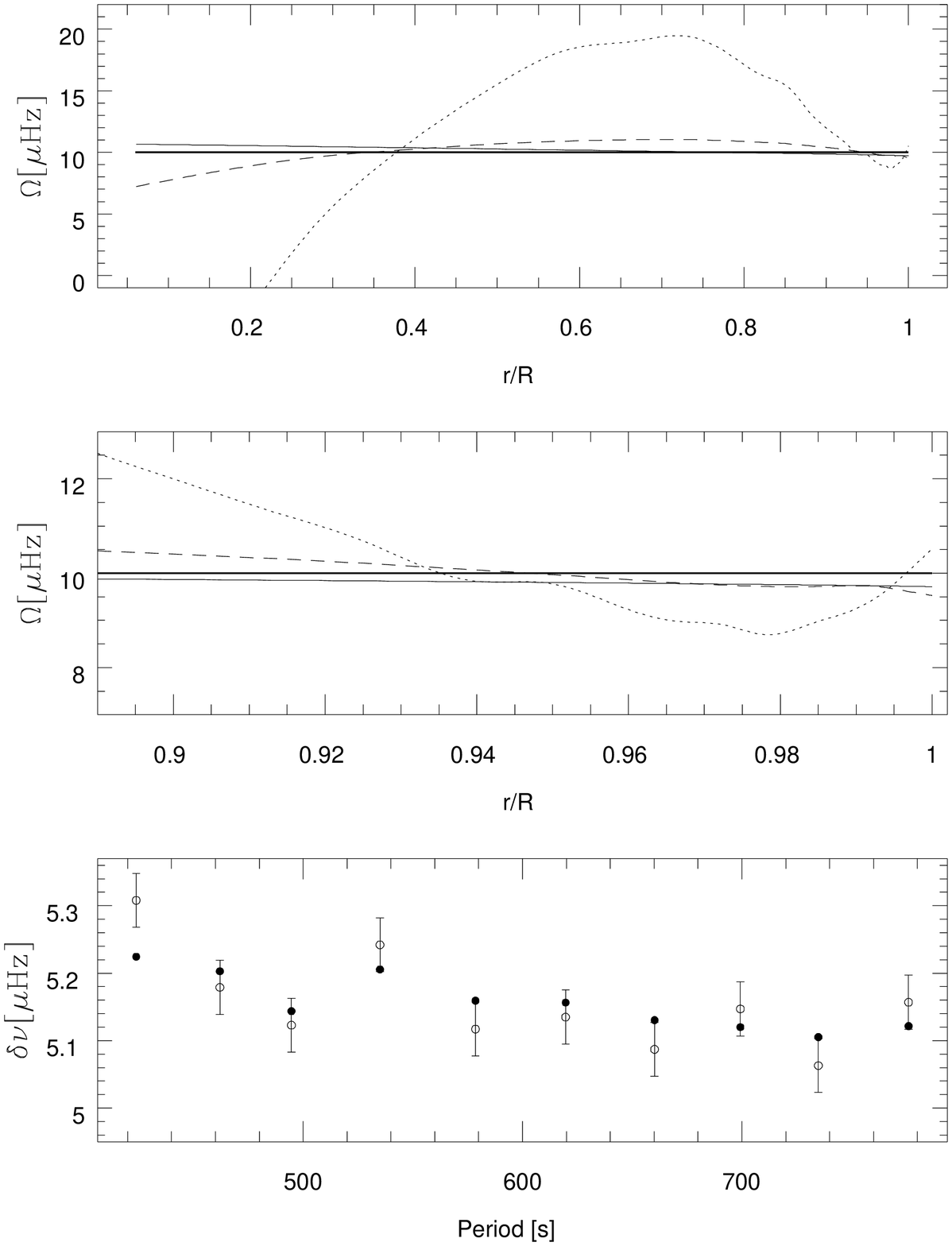]{Inversions for rotational splittings from a flat 
rotation curve in the GD~358 model (heavy solid line), but with errors 
in the splittings derived from a normal distribution.  The top panel 
shows the inversions for $\lambda=0.01$(dotted line), 0.1 (dashed 
line) and 1.0 (solid line) through the entire model; the middle panel 
shows an expansion of the radial scale for the outer layers.  The 
splittings used in the inversion are shown in the bottom panel. \label{fig13}}

%\plotone{fig14.ps}
\figcaption[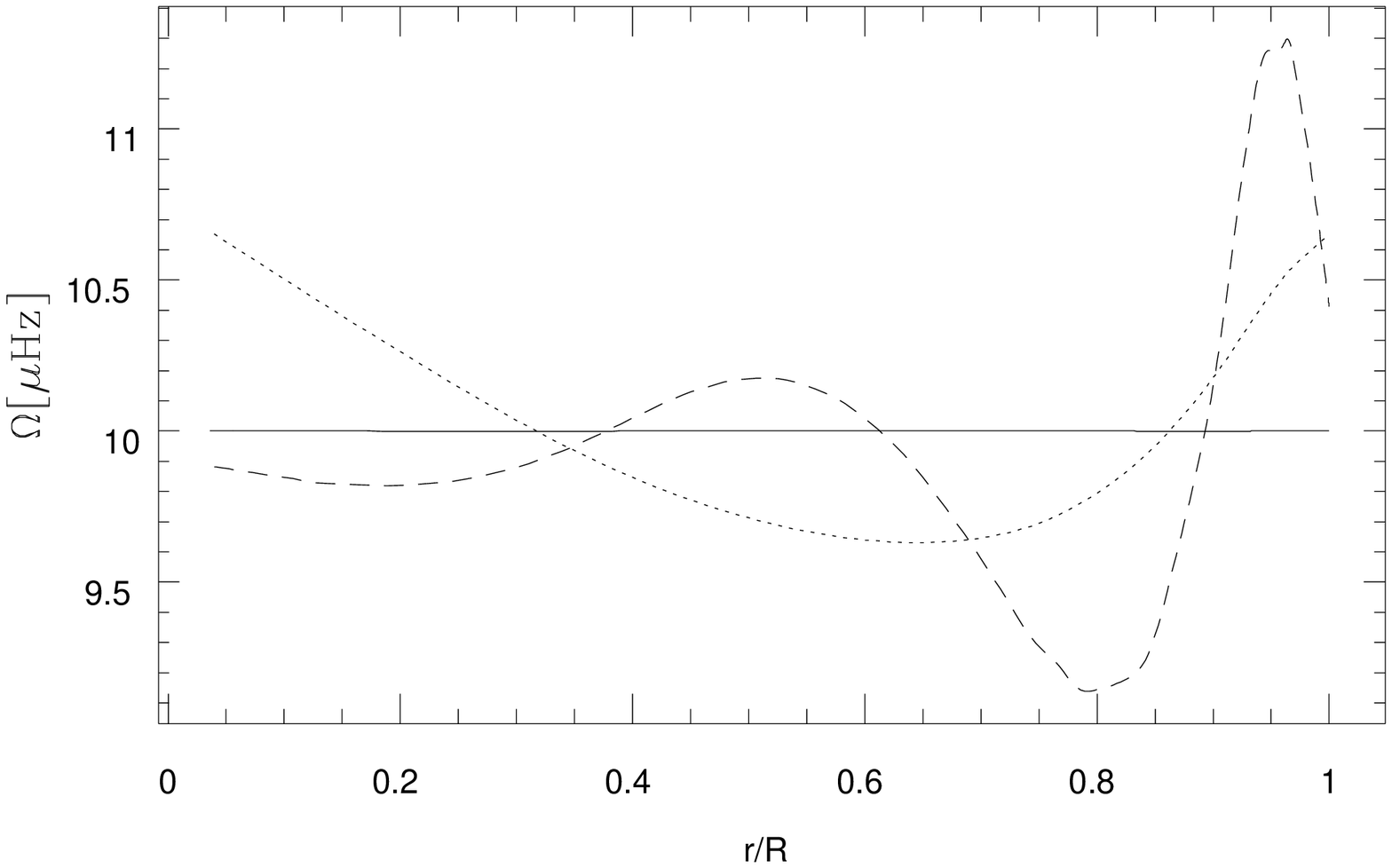]{Regularized inversions of noise-free splittings
of the PG~1159 model using ``incorrect'' kernels taken from a later model in
the evolutionary sequence.  The solid line shows the regularized inversion of
the correct model with $\lambda=0.01$; the dashed line is the inversion with
the incorrect model with the same value of $\lambda$.  The dotted line shows
the inversion with $\lambda=0.1$; there is no significant improvement in
increasing $\lambda$. \label{fig14}}

%\plotone{fig15.ps}
\figcaption[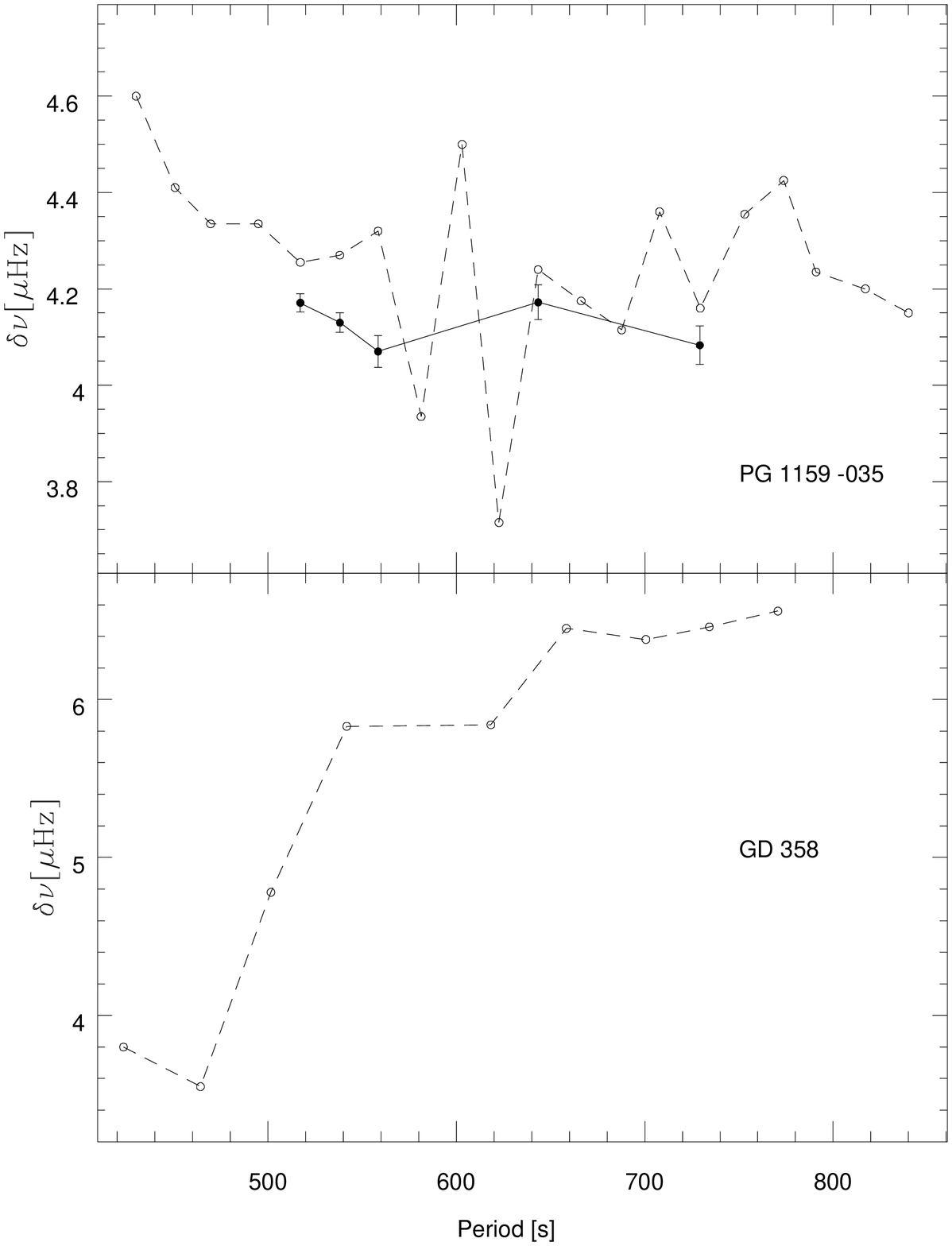]{Observed frequency splittings for PG~1159 (top
panel) and GD~358 (bottom panel).  The two sets for PG~1159 represent the
splittings derived in WWETPG from the power spectrum directly (open circles)
and by the least-squares fitting procedure described in the text (filled
circles).  The splittings for GD~358 are from the power spectrum analysis of
WWETDB. \label{fig15}}

%\plotone{fig16.ps}
\figcaption[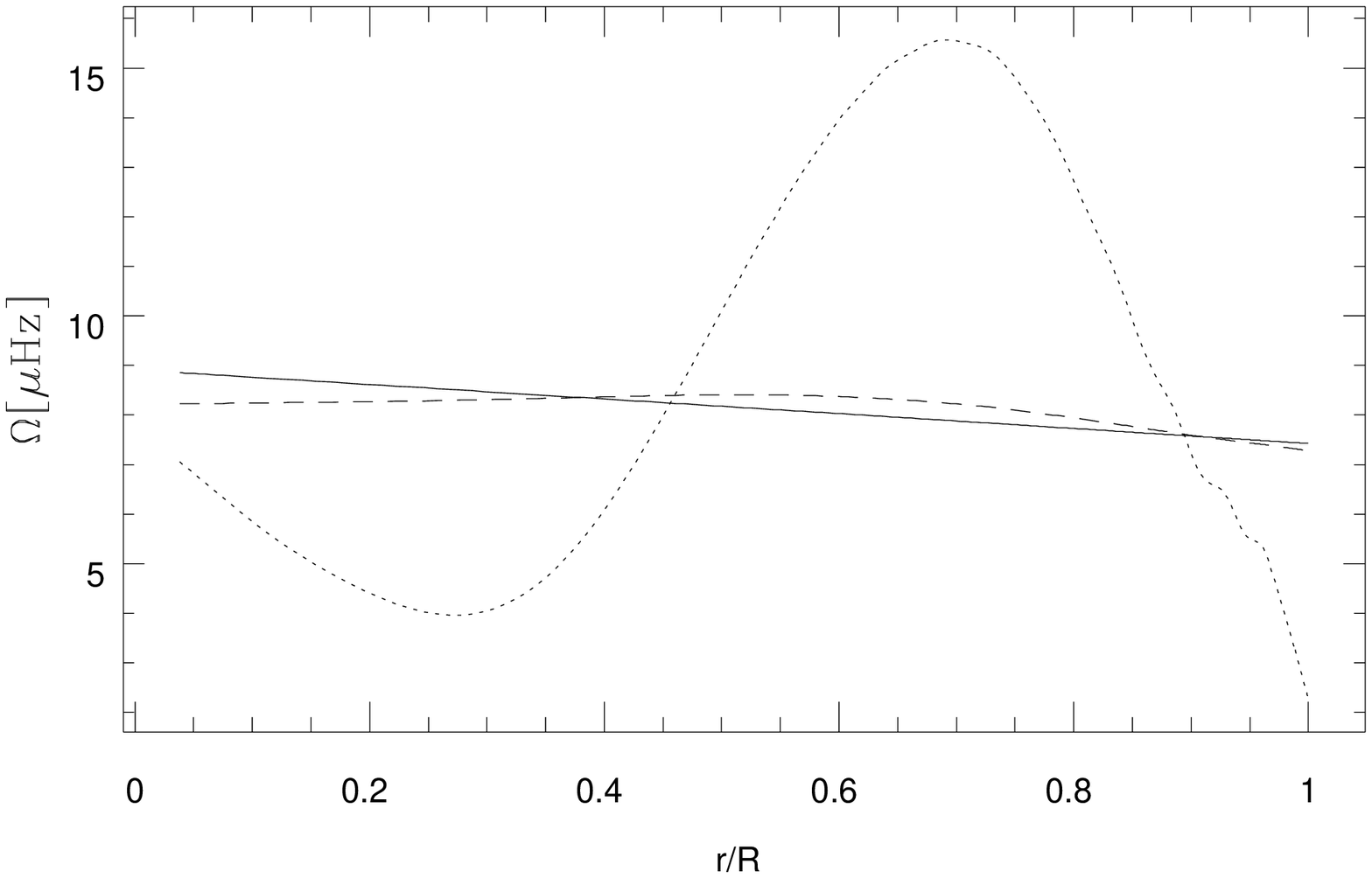]{Regularized inversions of the actual splittings
from PG~1159 for $\lambda=0.01$ (dotted line), 0.1 (dashed line), and 
1.0 (solid line). \label{fig16}}

%\plotone{fig17.ps}
\figcaption[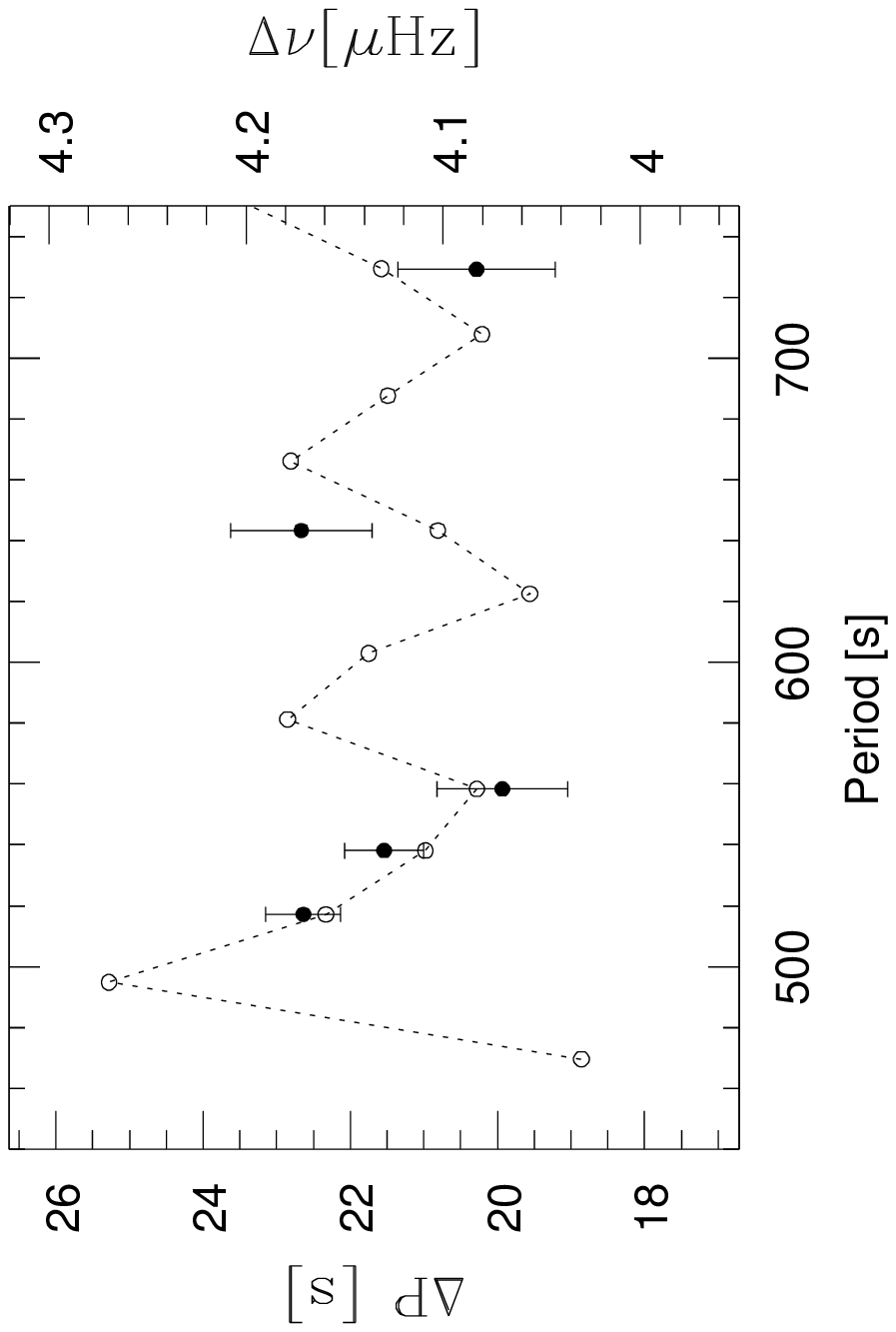]{Period spacing as a function of period for the
$l=1$ modes identified by WWETPG (dotted line, left-hand scale); data points
show the observed rotational splittings (right-hand scale).  Note that the
rotational splittings and period spacings change in phase.  This suggests
that the rotation rate decreases with radius in PG~1159. \label{fig17}}

%\plotone{fig18.ps}
\figcaption[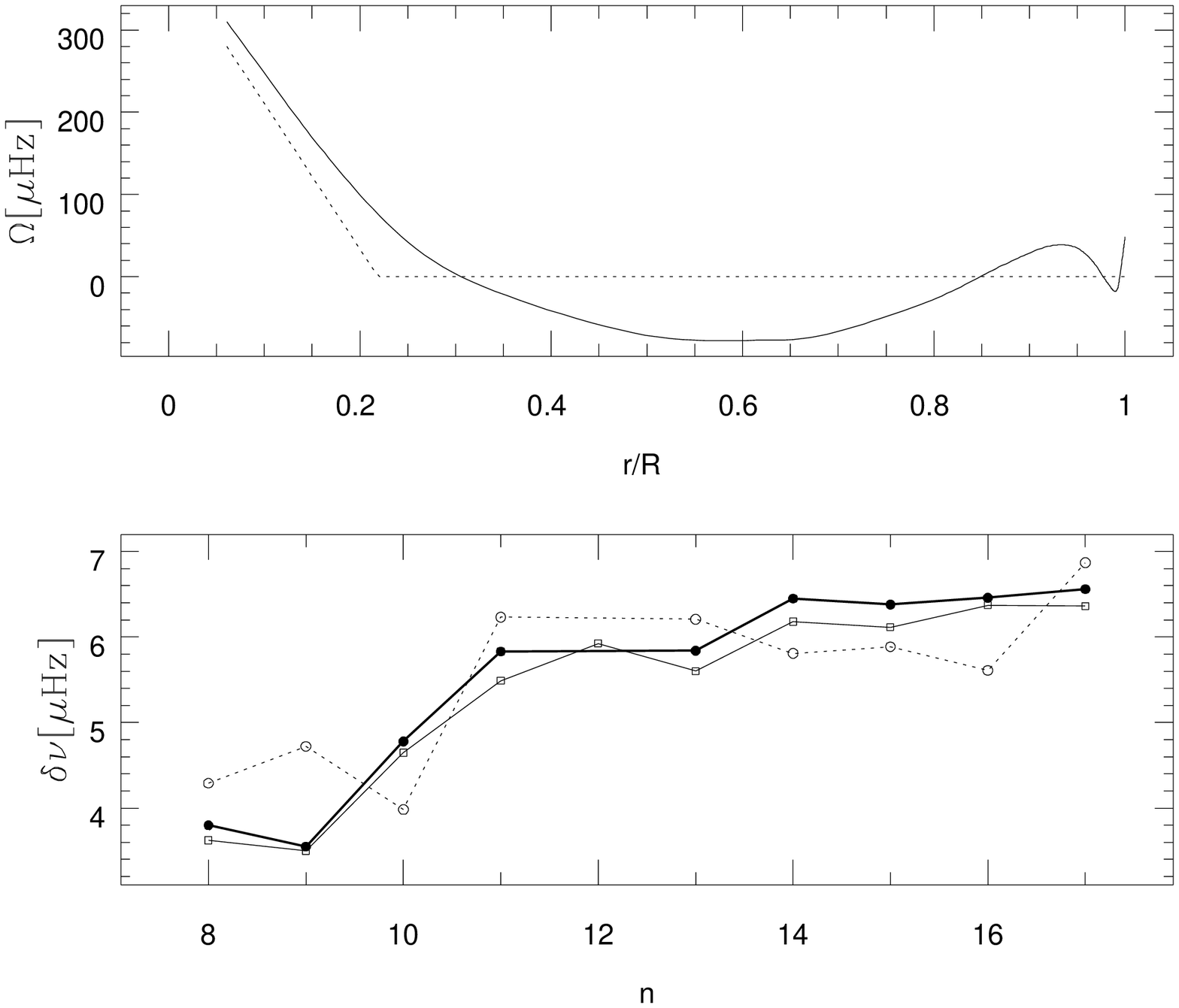]{Results of inversions of the observed frequency
splittings of GD~358.  The top panel shows the results of a regularized
inversion ($\lambda=10^{-2}$) as a solid line, and the results of
a function fitting (linear inside, flat outside) as a dotted line.
The bottom panel
shows the splittings computed for the given rotation curves.  The dark solid
line shows the observed splittings. \label{fig18}}

%\plotone{fig19.ps}
\figcaption[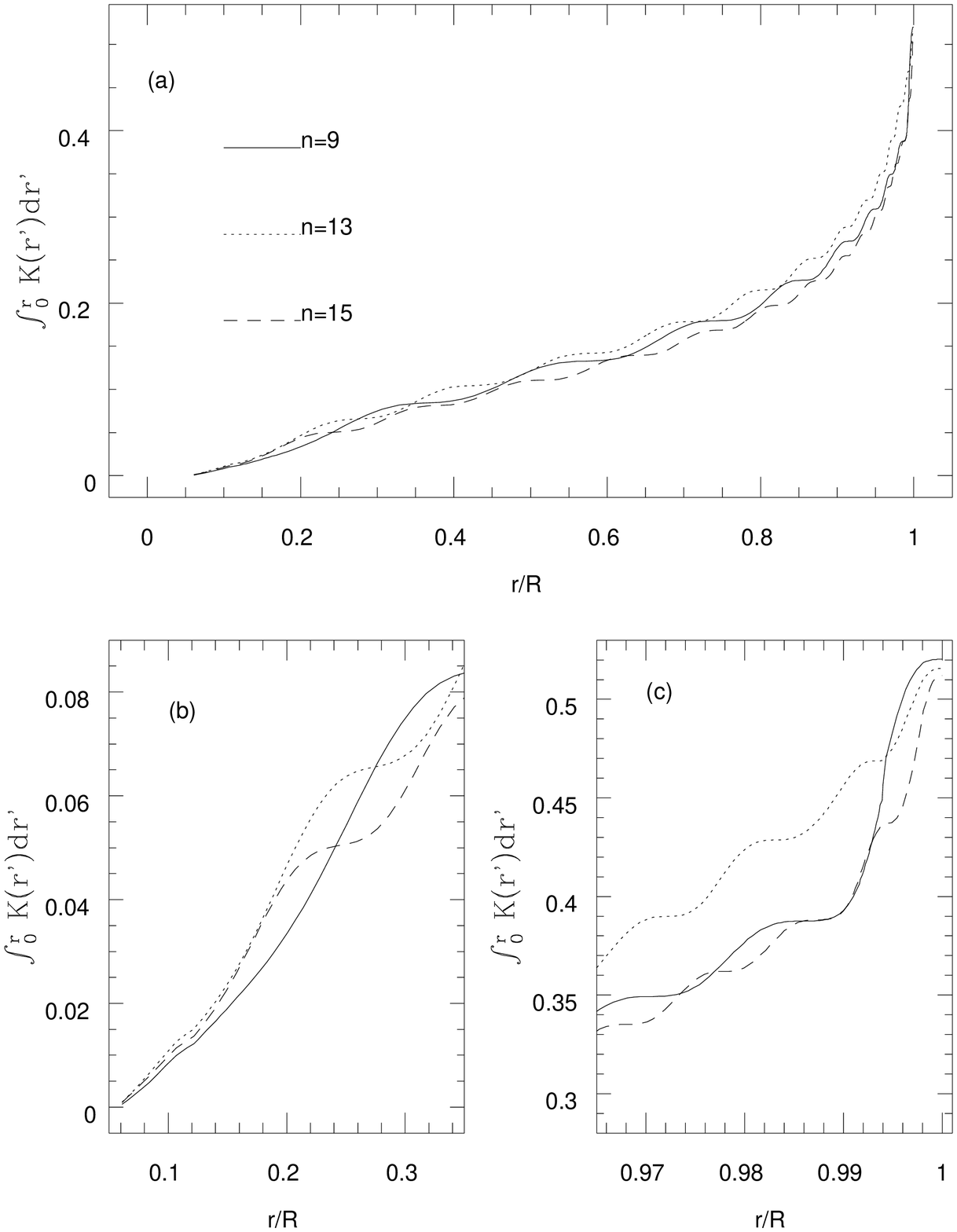]{The running integral of the rotation kernel for
selected modes in the GD~358 model.  The ordinate shows the value of the
integral up to the value at the abscissa; the value at $r/R$=1 corresponds to
$1-C_{n l}$.  Part (a) shows the integral over the entire interior, while part
(b) shows the central regions and part (c) shows the outer layers.
\label{fig19}}

%\plotone{fig20.ps}
\figcaption[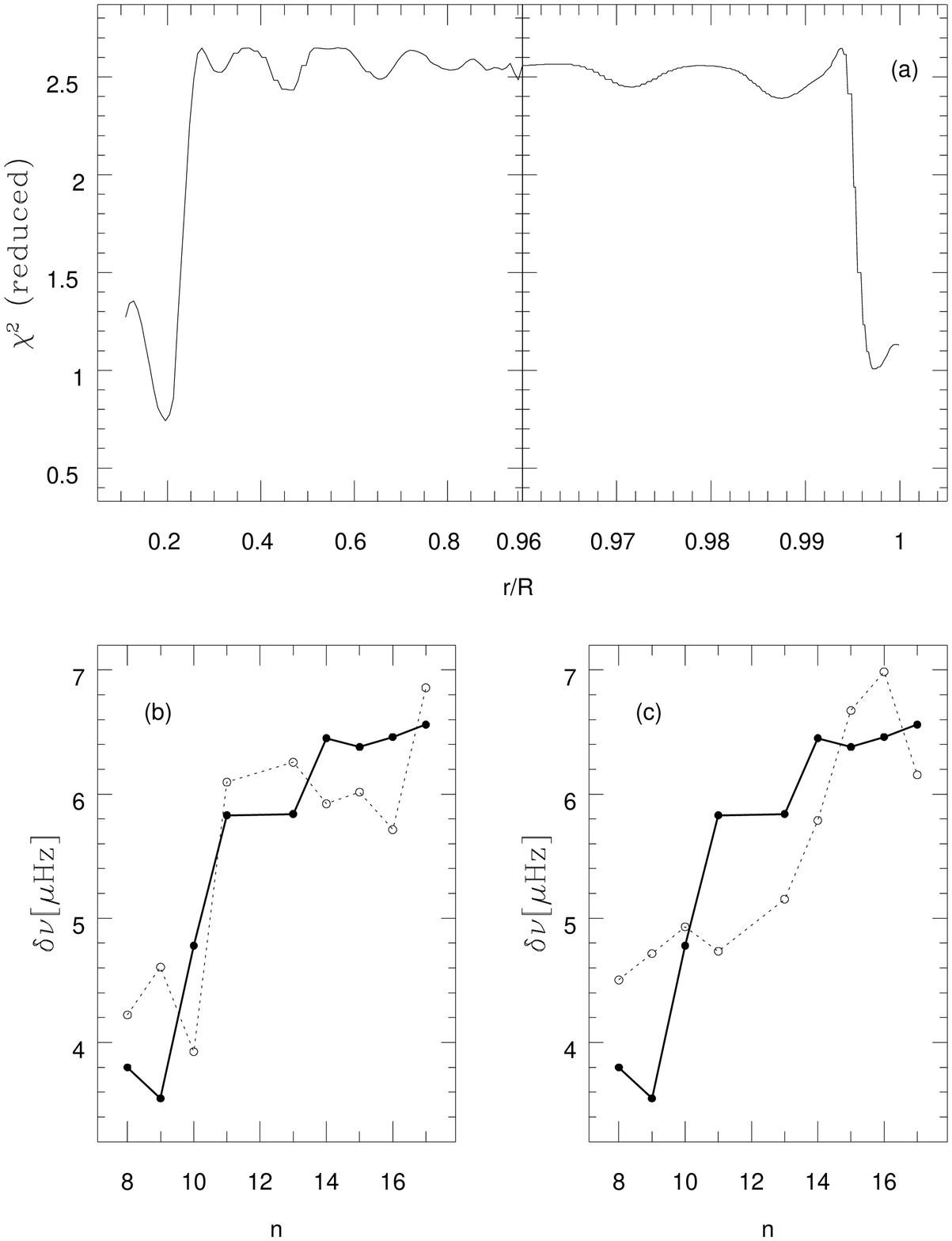] {(a) The run of $\chi^{2}$ for the function fitting
of the GD~358 splittings with a discontinuous rotation 
curve.  The X-axis shows the position of the discontinuity; the outer 
regions of the model are on an expanded horizontal scale.  The 
computed splittings with the inverted rotation curve (dotted line) are 
compared with the observed splittings (solid line) for the solution 
corresponding to the inner minimum of $\chi^{2}$ (b) and the outer 
minimum (c). \label{fig20}}

\clearpage

\begin{table}
\begin{center}
\caption{Frequencies and Splittings in PG 1159 \label{tab1}}
\vspace*{0.5in}

\begin{tabular}{rccc|ccc}
\multicolumn{4}{c}{WWETPG}&\multicolumn{3}{c}{Least Squares}\\
\tableline
n & Frequency & Amplitude & 2$<\delta \nu>$ & Frequency & Amplitude & 
2$<\delta \nu>$ \\
  & [\uhz]     &  [mma]     & [\uhz]    & [\uhz] & [mma] & [\uhz]   \\
\tableline
   & 1367.11 & 1.00 & & $1367.15 \pm 0.05$ & $ 0.94\pm 0.07$ \\
32 & 1370.80 & 0.56 & 8.32 & & & $8.17\pm 0.08$           \\
   & 1375.43 & 0.81 &  & $1375.32 \pm 0.06$ & $ 0.74\pm 0.07$ \\
 & & & \\
   & 1550.39 & 1.07 & & $1550.43 \pm 0.05$ & $ 0.87\pm 0.07$ \\
28 & 1554.23 & 0.55 & 8.49 & & & $8.34\pm 0.07$ \\
   & 1558.88 & 1.18 & & $1558.77 \pm 0.05$ & $ 1.00\pm 0.07$ \\
 & & & \\
   & 1786.24 & 1.29 & & $1786.70 \pm 0.06$ & $ 0.85\pm 0.09$ \\
24 & 1790.70 & 2.46 & 8.64 & $1790.67 \pm 0.02$ & $ 2.53\pm 0.08$ & $8.14\pm 0.07$ \\
   & 1794.88 & 3.30 & & $1794.84 \pm 0.02$ & $ 3.01\pm 0.09$ \\
 & & & \\
   & 1854.04 & 6.10 & & $1854.06 \pm 0.01$ & $ 5.43\pm 0.09$ \\
23 & 1858.20 & 4.16 & 8.54 & $1858.20 \pm 0.01$ & $ 3.82\pm 0.08$ & $8.26\pm
0.04$ \\
   & 1862.58 & 2.73 & & $1862.32 \pm 0.04$ & $ 1.45\pm 0.09$ \\
 & & & \\
   & 1929.32 & 5.09 & & $1929.44 \pm 0.02$ & $ 3.57\pm 0.09$ \\
22 & 1933.55 & 4.25 & 8.51 & $1933.61 \pm 0.01$ & $ 3.87\pm 0.08$ &$8.34\pm
0.02$ \\
   & 1937.83 & 6.87 & & $1937.78 \pm 0.01$ & $ 6.27\pm 0.09$ 
\end{tabular}
\end{center}
\end{table}

\end{document}